\documentclass[aps,twocolumn,showpacs,superscriptaddress,prb,10pt]{revtex4-1}
\usepackage{graphicx}
\usepackage{dcolumn}
\usepackage{bm}
\usepackage{color}
\usepackage{amsmath}
\usepackage{amsfonts}
\usepackage{subfigure}
\usepackage[makeroom]{cancel}

\newcommand{\COMMENTED}[1]{}

\begin{document}

\author{Ettore Vitali}
\affiliation{Department of Physics, The College of William and Mary, Williamsburg, Virginia 23187}

\author{Hao Shi}
\affiliation{Department of Physics, The College of William and Mary, Williamsburg, Virginia 23187}

\author{Mingpu Qin}
\affiliation{Department of Physics, The College of William and Mary, Williamsburg, Virginia 23187}

\author{Shiwei Zhang}
\affiliation{Department of Physics, The College of William and Mary, Williamsburg, Virginia 23187}

\title{Computation of dynamical correlation functions 
for many fermion systems with auxiliary-field quantum Monte Carlo} 

\begin{abstract}
We address the
  calculation of dynamical correlation functions
for many fermion systems at zero temperature,
using the auxiliary-field quantum Monte Carlo method.
The two-dimensional Hubbard hamiltonian is used as a model system.
Although most of the calculations performed here are for cases where the sign problem is absent,
the discussions are kept general for applications to physical
problems when the sign problem does arise.
We study the use of 
twisted boundary conditions to improve the extrapolation of the results to the thermodynamic limit.
A strategy is proposed to drastically reduce finite size effects relying on a minimization
among the twist angles. 
This approach is demonstrated by computing
the charge gap at half-filling.
We obtain accurate results showing the scaling of the
gap with the interaction strength $U$, connecting to the
scaling of the unrestricted Hartree-Fock method at small $U$ and Bethe
Ansatz exact result in one dimension at large $U$.
\COMMENTED{In the weakly interacting regime, the dependence of
the charge gap on the interaction strength
is captured by unrestricted Hartree-Fock theory with renormalized
parameters, while, for larger values of the interaction, a behavior
consistent with the Bethe ansatz exact solution in one--dimension is observed. }
A new algorithm is then proposed to compute dynamical Green functions and correlation functions
which explicitly varies the number of particles during
the random walks in the manifold of Slater determinants.
In dilute systems, such as ultracold Fermi gases, 
this algorithm enables
calculations with much more favorable complexity, 
with computational cost proportional to basis size or the number of lattice sites.
\end{abstract}

\maketitle

\section{Introduction}
The calculation of dynamical correlation functions of 
many-body quantum systems 
is
a great challenge in theoretical condensed matter physics.
Such functions provide 
a unique
opportunity to explore the manifold of the excited states of a 
physical system.
They often provide a much more direct connection to experimental measurements, 
 giving access to 
crucial properties 
such as spectral
functions, excitation spectra, and charge and spin gaps, to name a few.

With the advent of modern computational resources, 
quantum Monte Carlo (QMC) simulations \cite{foulkes, AFQMC-lecture-notes-2013,Assaad-lecture-notes-2002,Alavi} 
are becoming a very powerful tool for 
 computations in quantum 
many-particle models and realistic systems. 
A vast array of total energy and related quantities have been computed.
Equal-time correlation functions have also been studied fairly routinely by QMC, especially in 
model systems.
To estimate dynamical properties
from first principles using QMC has been more challenging.
Important results have been obtained for
bosonic strongly correlated systems.
\cite{gift,overpressurized,rota_2013,nava_superfluid_2012,saccani_bose_2011,saccani_excitation_2012,Molinelli2016}
Also excited states of low dimensional
bosonic systems have been recently studied \cite{arrigoni_excitation_2013,helium1D}. For fermionic systems, 
in special situations where the sign problem is not present, accurate calculations have been performed \cite{PhysRevB.31.4403,assaad_prb,
assaad_prl, assaad_recent}.  
A few attempts have been also made \cite{PhysRevB.31.4403,Booth,he3_dyn, jel_dyn_method, jel_dyn_method2}
in the more difficult situations
where approximations are needed to deal with the sign problem \cite{FNKalosJcp, DMC, RevModPhys.73.33, cpprl1995, phaseless2003}. 

In this paper we study the calculation of imaginary-time
correlation functions and excited state information in interacting
fermion systems at zero temperature. 
We formulate and discuss our methods in a general framework,
keeping in mind applications in 
many-fermion systems where a sign problem arises and a  
constrained path approach is applied within an open-ended imaginary-time
projection. 
Most of our illustrations and applications in the present paper, however, will be in
systems where the sign problem is absent and exact results can be obtained.
For these we use the 
 two-dimensional Hubbard hamiltonian 
with repulsive interaction at half-filling.
We 
employ the path-integral auxiliary-field quantum Monte Carlo (AFQMC) framework,
but including a force-bias acceleration technique \cite{Hao-2DFG} in the Metropolis sampling and a technique 
to control Monte Carlo variance divergence \cite{Hao-inf-var}.
\COMMENTED{in systems where the sign problem is absent. However, the
discussions are kept general for applications within a
constrained path approach of open-ended imaginary-time
projection, when the sign problem does arise.
We use the 
 two-dimensional Hubbard hamiltonian 
as a model system. }

In the first part of the paper, 
we implement a standard approach \cite{assaad_prb,hirsch-stable} of computing non-equal-time Green functions 
and correlation functions within AFQMC.
\COMMENTED{hirsch-stable,
the calculation of dynamical
correlations following the
methodology suggested by Assaad et al. \cite{...} 
relying on a very efficient Auxiliary Field Quantum Monte Carlo
algorithm exploiting a force bias to improve the 
sampling of auxiliary fields.}
Our focus is on the extrapolation of the results
to the thermodynamic limit and testing the efficiency of different implementations in general many-fermion 
systems. 
We show
that it is convenient to introduce twisted boundary
conditions \cite{PhysRevE.64.016702}, 
and suggest a way to exploit the boundary conditions
that dramatically reduces finite-size effects 
in the calculation of the charge gap. 
Accurate results for the gap are obtained  in the repulsive Hubbard model 
even at weak interactions,
which mimic many real materials where the gap might be very small 
compared to the energy scales, presenting challenges for numerical calculations.
With our approach, the charge gap is determined even at $U/t$ as small as $0.5$,
far beyond the reach of previous unbiased many-body calculations.
\COMMENTED{
We succeed in reaching an 
accuracy that allows to detect a non-zero charge gap
for the repulsive two-dimensional Hubbard model even
at $U/t$ as small as $0.5$.
}

In the second part, 
we propose a new algorithm to compute
dynamical Green functions, density-density and spin-spin correlation functions which, in the dilute limit, dramatically reduces 
the complexity without
affecting the numerical stability of the calculations or the
accuracy of the results.
The method relies on the explicit variation of the number of particles during
the random walk in the manifold of Slater determinants.
 We show that, for example, the spectral function for one given momentum can be calculated
with a 
computational complexity
proportional to the number of lattice
sites, which enables 
simulations on very large lattices. 
\COMMENTED{We demonstrate this algorithm in the atomic Fermi gas, 
which can be modeled by the attractive Hubbard model 
in the dilute limit \cite{PhysRevA.86.013626}. 
Results for lattices with over 
$2000$ sites 
are obtained to provide an estimate of the pairing gap
for a two-dimensional Fermi gas in the strongly interacting BEC-BCS crossover regime. 
}
The new method will have great advantages in systems such as atomic Fermi gases,
which can be modeled by the 
attractive Hubbard model in the dilute limit \cite{PhysRevA.86.013626},
or real materials, where the calculations typically require \cite{phaseless2003}
a basis size much larger than the
number of electrons. 

The reminder of this paper is organized as follows. In Sec.~II we will briefly sketch the AFQMC 
methodology, 
and then describe our implementation of a stable algorithm to compute dynamical
Green functions, 
both in the path-integral (for sign-problem-free systems) and open-ended random walk (for 
constrained path calculations when the sign problem is present).
In Sec.~III, we propose a strategy to control finite-size effects using
twisted boundary conditions, 
and present our results
for the charge gap in the repulsive Hubbard model at half-filling
in the intermediate and weakly  interacting regime.
In Sec.~IV 
we describe our new approach which, in the dilute limit, dramatically improves the complexity of the calculations. 
We conclude in Sec.~V.



\section{Auxiliary Field Quantum Monte Carlo formalism}

\COMMENTED{
The Path Integral Auxiliary Field Quantum Monte Carlo (AFQMC) is a zero temperature projector Quantum
Monte Carlo method. Since we use the Hubbard hamiltonian
as a model system, we introduce the basic notations of the
methodology using, as a starting point, the hamiltonian:
}

We introduce the basic notations of the
methodology using the Hubbard hamiltonian which, 
as mentioned, will be our model system:
\begin{equation}
\label{ham}
\hat{H} =  -t \sum_{<i,j>, \sigma = \uparrow,\downarrow} \hat{c}^{\dagger}_{i,\sigma} \hat{c}^{}_{j,\sigma} + U \sum_{i} \left( \hat{n}_{i,\uparrow} - \frac{1}{2}\right)  \left(\hat{n}_{i,\downarrow} - \frac{1}{2}\right)
\end{equation}
where the labels $i,j$ run over the sites of a square lattice with $\mathcal{N}_s = L_x \times L_y=L^2$ sites,
the symbol $<,>$ denotes, as usual, nearest neighbors and $\hat{n}_{i,\sigma} = \hat{c}^{\dagger}_{i,\sigma} \hat{c}^{}_{i,\sigma}$ is the particle number density operator on site $i$ for the given spin direction $\sigma$.
The Hubbard model\cite{Hubbard_1,Hubbard_2}  is one of the most
widely studied models in condensed matter physics.
It is of fundamental theoretical importance in the context
of magnetism and cuprate superconductors, and is relevant to experiments using ultracold
fermionic atoms both in a trap (in the continuum) and in optical lattices. 
Despite its simplicity, no analytical solutions to this
Hamiltonian are known beyond the perturbative limit.
The model provides
an excellent 
test ground for many-body theories and
computational approaches.
\COMMENTED{
 and, besides its well known
ability to describe interaction driven metal-insulator
transition and magnetism, has been shown to capture 
physical properties of high-$T_c$ superconductors (in the
repulsive case), and to provide a regularization of the hamiltonian
of a dilute Fermi atomic cold gas (in the attractive case) .
 Although this model has been introduced
several years ago, its comprehensive solution still remains
a grand challenge in condensed matter physics.
}

Denoting by $|\phi_T\rangle$ a Slater determinant with $N_{\uparrow}$ spin-up and
 $N_{\downarrow}$  spin-down particles, provided that $|\phi_T\rangle$ is not orthogonal to the $\mathcal{N}_p$-particle ($\mathcal{N}_p=N_{\uparrow}+N_{\downarrow}$) ground state $ | \,\Psi_0 \, \rangle$ of \eqref{ham}, the following relation holds:
 \begin{equation}
| \, \Psi_0 \rangle \propto \lim_{\beta \to +\infty} e^{-\beta ( \hat{H} - E_0)} |\phi_T\rangle
\end{equation}
where $E_0$ is an estimate of the ground state energy.
A combined use of Trotter-Suzuki breakup and Hubbard-Stratonovich transformation provides
the following approximation:
 \begin{equation}
 \label{propagator}
e^{-\beta ( \hat{H} - E_0)}  = \left(e^{-\delta\tau ( \hat{H} - E_0)} \right)^{M} \simeq  \left(\int d{\bf{x}} p({\bf{x}}) \hat{B}({\bf{x}})\right)^{M} 
\end{equation}
where ${\bf{x}} = (x_1, \dots, x_{\mathcal{N}_s})$ is an auxiliary field (often discrete Ising fields on the lattice),
$\hat{B}({\bf{x}})$ is a one-particle propagator, and $\delta\tau = \beta/M$
is a sufficiently small {\it{time-step}}. The function $p({\bf{x}})$ is a 
probability density.

In the repulsive Hubbard model, for example, the simplest way to build the approximation in Eq.~\eqref{propagator}
is to use the following discrete spin decomposition of the Hubbard-Stratonovich transformation:
\begin{equation}
 \label{hs}
e^{-\delta \tau U \hat{n}_{i,\uparrow}\hat{n}_{i,\downarrow} } = 
\sum_{x_i = \pm 1} \frac{1}{2} \hat{b}_i(x_i)
\end{equation}
where:
\begin{equation}
 \label{hs-b}
\hat{b}_i(x) = e^{-\delta \tau U \frac{\hat{n}_{i,\uparrow} + \hat{n}_{i,\downarrow}}{2}}
\, e^{-\gamma x (\hat{n}_{i,\uparrow} - \hat{n}_{i,\downarrow}) } 
\end{equation}
where $\cosh(\gamma)=\exp(\delta \tau U/2)$.
Different forms of the decomposition can affect the accuracy and efficiency of the calculation \cite{PhysRevB.88.125132,mingpu-to-be-published}, but will not affect the formalism discussed in the rest of the paper.

A key point of the methodology is that the operator $\hat{B}(\bf{x})$
appearing in  Eq.~\eqref{propagator} is the exponential of a one-body operator dependent on the auxiliary field
configuration. We can write:
\begin{equation}
\label{b_operator}
\hat{B}(\bf{x}) = \exp\left(\sum_{i\sigma,j\sigma'} \mathcal{A}_{i\sigma,j\sigma'}(\bf{x}) \,\,\,
\hat{c}^{\dagger}_{i,\sigma} \,\hat{c}^{}_{j,\sigma'} \right)
\end{equation}
where the explicit form of the $\mathcal{N}_s \times \mathcal{N}_s$-matrix $ \mathcal{A}(\bf{x})$ depends
on the choice of the Hubbard-Stratonovich transformation.
For the description of the formalism, it will turn out to be useful to introduce the
matrix:
\begin{equation}
\label{b_matrix}
\mathcal{B}(\bf{x}) = \exp\left(\mathcal{A}(\bf{x})\right)\,.
\end{equation}
Any operator $\hat{B}(\bf{x})$ operating on a Slater determinant $|\phi\rangle$ results in 
another Slater determinant $|\phi'\rangle$, given in matrix form by
\begin{equation}
\mathcal{B}(\bf{x}) \Phi =\Phi'\,,
 \label{eq:Thouless}
\end{equation}
where 
$\Phi=\Phi_\uparrow\otimes \Phi_\downarrow$, with $\Phi_\sigma$ being the $\mathcal{N}_s\times N_\sigma$ matrix 
containing the spin-$\sigma$ orbitals of the Slater determinant wave function $|\phi\rangle$,
and similarly for $|\phi'\rangle$.

\subsection{Static properties}
The standard path-integral AFQMC method allows the evaluation of ground state
expectation values:
\begin{equation}
\label{exp_val}
\langle \hat{O} \rangle = \frac{ \langle \Psi_0  | \, \hat{O} \, | \Psi_0 \rangle }
{\langle \Psi_0 \, | \, \Psi_0 \rangle }
\end{equation}
by casting them in the following form:
\begin{equation}
\label{exp_val2}
\langle \hat{O} \rangle = \int d{\bf{X}} \, \mathcal{W}({\bf{X}}) \, \mathcal{O}({\bf{X}})\,.
\end{equation}
In Eq.~\eqref{exp_val2}, ${\bf X} = ({\bf x}(1), \dots, {\bf x}(M))$ denotes a (discretized) {\it{path}} in auxiliary
fields configurational space. Moreover, if we introduce the 
two Slater Determinants:
\begin{equation}
\label{phi_L}
\langle \phi_L  | \,  = \langle \phi_T  | \,\hat{B}({\bf x}(M)) \dots \hat{B}({\bf x}(l)) 
\end{equation}
and:
\begin{equation}
\label{phi_R}
| \phi_R \rangle = 
\hat{B}({\bf x}(l-1)) \dots \hat{B}({\bf x}(1))  \,| \phi_T \rangle\,,
\end{equation}
we may write:
\begin{equation}
 \mathcal{W}({\bf{X}}) \propto \, \langle \phi_L \, |   \, \phi_R \rangle \,  \prod_{i=1}^{M} p({\bf{x}}(i))
\end{equation}
and:
\begin{equation}
\label{static_estimator}
\mathcal{O}({\bf{X}}) = \frac{ \langle \phi_L  | \, \hat{O} \, | \phi_R \rangle }
{\langle \phi_L \, | \, \phi_R \rangle }
\end{equation}
Whenever $ \mathcal{W}({\bf{X}}) \geq 0$ for each auxiliary field configurations,
as it happens if $U<0$ and $N_\uparrow=N_\downarrow$ or at half-filling ($\mathcal{N}_p=\mathcal{N}_s$) when $U>0$, the integral in \eqref{exp_val2} can
be evaluated via Monte Carlo. 
We use an efficient Metropolis sampling of the paths, exploiting a force 
bias \cite{Hao-2DFG,AFQMC-lecture-notes-2013} that allows high acceptance ratio in the updates of the
path in the auxiliary field configuration space, and eliminating the infinite variance problem \cite{Hao-inf-var}
with a bridge link approach.

\COMMENTED{
A key point of the methodology is that the operator $\hat{B}(\bf{x})$
appearing in  \eqref{propagator} is a one-particle propagator dependent on the auxiliary field
configuration. We can write:
\begin{equation}
\label{b_operator}
\hat{B}(\bf{x}) = \exp\left(\sum_{i\sigma,j\sigma'} \mathcal{A}_{i\sigma,j\sigma'}(\bf{x}) \,\,\,
\hat{c}^{\dagger}_{i,\sigma} \,\hat{c}^{}_{j,\sigma'} \right)
\end{equation}
where the explicit form of the matrix $ \mathcal{A}(\bf{x})$ depends
on the choice of the Hubbard-Stratonovich transformation.
For the description of the formalism, it will turn out to be useful to introduce the
matrix:
\begin{equation}
\label{b_matrix}
\mathcal{B}(\bf{x}) = \exp\left(\mathcal{A}(\bf{x})\right)
\end{equation}
}

When a sign problem is present, a constrained path \cite{cpprl1995} or phase-free approximation \cite{phaseless2003} can be imposed to 
remove the exponentially growing noise (with system size or the length of the path $\beta$) and restore
the same computational scaling as in the sign-problem-free cases. The paths become open-ended. 
The primary difference is that $\langle \phi_L|$ is now produced by back-propagation (BP) \cite{AFQMC-lecture-notes-2013}, and a weight 
accompanies each path. 
The details of the back-propagation for static properties 
have been discussed elsewhere \cite{Wirawan-boson}. The modification to compute dynamical
properties is minimal beyond what is necessary for the path-integral formalism, and we will comment on 
it as needed below.

\subsection{Dynamical properties}
\label{ssec:method-dynamical}

Dynamical correlation functions in imaginary-time
at zero temperature have
the general form:
\begin{equation}
\label{dynamical}
f(\tau) = \frac{ \langle \Psi_0  | \, \hat{A} \, e^{-\tau (\hat{H} - E_0) } \, \hat{B} \, | \Psi_0 \rangle }
{\langle \Psi_0 \, |  \, \Psi_0 \rangle }
\end{equation}
where $\hat{A}$ and $\hat{B}$ can be destruction or creation operators, or one-body
operators such as the particle density or the spin density or even more general
operators.

\COMMENTED{ 
An efficient way to calculate \eqref{dynamical} in the context
of AFQMC has been introduced by Assaad et al. \cite{???},
and we implemented a similar algorithm in our AFQMC code.
We are going now to briefly review the calculation strategy.
Incidentally, we observe that the formalism we will use, although specific of the
Path-Integral method, can be easily modified to be used within a constrained
path approach of open-ended imaginary time projection; this becomes necessary
when the sign problem is present. More precisely, a stochastic sampling of
Slater determinants defined as in \eqref{phi_R} is performed, and provides
{\it{mixed}}  estimators of the form $\langle \phi_T | \hat{O} \, | \Psi_0 \rangle$,
$| \phi_T \rangle$ playing the role of starting point of the open-ended random walk.
In order to get {\it{pure}} estimators, $\langle \Psi_0 | \hat{O} \, | \Psi_0 \rangle$, a
back-propagation strategy is used, meaning that $| \phi_T \rangle$  is propagated
backword, building a bra of the form \eqref{phi_L}. Naturally, an approximate
constraint on the random walk in the manifold of Slater Determinants is
needed to avoid the signal to noise ratio becoming exponentially small 
with the number of particles. Keeping this in mind, all the formalism
we will describe can be easily adapted to any AFQMC methodology.
}

Let us focus on the dynamical {\it{particles}} and {\it{holes}} 
Green functions
in imaginary-time, i.e. the matrices:
\begin{equation}
\label{particles}
{{G^{p}}}_{i\sigma,j\sigma'}(\tau) =   \frac{ \langle \Psi_0  | \, \hat{c}^{}_{i,\sigma} \, e^{-\tau  (\hat{H} - E_0) } \,  \hat{c}^{\dagger}_{j,\sigma'}  \, | \Psi_0 \rangle }
{\langle \Psi_0 \, | \, \Psi_0 \rangle }
\end{equation}
 and:
\begin{equation}
\label{holes}
{{G^{h}}}_{i\sigma,j\sigma'} (\tau) = \frac{ \langle \Psi_0  | \, \hat{c}^{\dagger}_{i,\sigma} \, e^{-\tau ( \hat{H} - E_0)} \,  \hat{c}^{}_{j,\sigma'} \, | \Psi_0 \rangle }
{\langle \Psi_0 \, |  \, \Psi_0 \rangle }\,. 
\end{equation}
When the Hamiltonian is spin-independent as is the case in the Hubbard model, all the terms with $\sigma \neq \sigma'$ identically vanish.
In order to keep the notations simple, we will work for $\sigma = \, \uparrow$ (the other
case being analogous) and
neglect the spin index.
Dealing with translationally invariant systems, 
we will denote 
${{G^{p,h}}}_{i\sigma,j\sigma}(\tau)$ by $G^{p,h}(R,\tau)$ with $R = (i - j)$. The Fourier transforms, 
i.e., the dynamical Green functions in momentum space, will be denoted by $\tilde{G}^{p,h}(Q,\tau)$,
where 
$Q=(Q_x,Q_y)$ is a wave-vector of the
reciprocal lattice:
$Q_{x} = \frac{2\pi}{L_x}\, n_{x}$, with $n_{x} \in \{0, \dots, L_x-1\}$,
and similarly for $Q_{y}$.

The imaginary-time propagator between the operators  $\hat{A}$ and $\hat{B}$ 
can again be expressed using Eq.~\eqref{propagator}. 
This can be thought of as inserting an extra segment to the path we keep:
a number $N_{\tau} = \tau/\delta\tau$ of
{\it{time-slices}}, say ${\bf \tilde{x}}(1), \dots , {\bf \tilde{x}}(N_{\tau})$.
The static estimator Eq.~\eqref{static_estimator} is replaced 
by a dynamical estimator which (for example in the case of the particles Green function) 
can be cast in the form (see Ref.~\onlinecite{PhysRevB.31.4403}):
\begin{equation}
\label{dynamical estimator}
g({\bf{X}},\tau) = \frac{ \langle \phi_L  | \,  \hat{c}^{}_{i} \, \hat{B}({\bf{\tilde{x}}}(N_{\tau})) \dots  \hat{B}({\bf{\tilde{x}}}(1))  \,\,\hat{c}^{\dagger}_{j}  | \phi_R \rangle }
{\langle \phi_L \, |  \hat{B}({\bf{\tilde{x}}}(N_{\tau})) \dots  \hat{B}({\bf{\tilde{x}}}(1)) \,| \, \phi_R \rangle }
\end{equation}
 To keep the notation simple, we will write $\hat{B}_i$ instead of  $\hat{B}({\bf \tilde{x}}(i))$
 from now on.

In order to calculate Eq.~\eqref{dynamical estimator} for a given configuration
of auxiliary fields, we use the manipulations presented
in Ref.~\onlinecite{assaad_prb}. 
We introduce the equal-time Green function matrix:
\begin{equation}
 \mathcal{G}^p(n, n) = \frac{\langle \phi_L \, | \, \hat{B}_{N_{\tau}} \dots \hat{B}_{n+1} \, \hat{c}^{}_{i} \,  \hat{c}^{\dagger}_{j} \,  \hat{B}_n\hat{B}_{n-1} \dots \, \, \hat{B}_1 | \phi_R \rangle} 
 {\langle \phi_L \, |  \,\,  \hat{B}_{N_{\tau}} \dots  \hat{B}_1 \,\, | \phi_R \rangle}
\end{equation}
and the time displaced one:
\begin{equation}
\begin{split}
& \mathcal{G}^p(n, m)=  \frac{\langle \phi_L \, | \, \hat{B}_{N_{\tau}} .. \hat{B}_{n+1} \, \hat{c}^{}_{i} \,\,  \hat{B}_n ..  \hat{B}_{m+1} \,\, \hat{c}^{\dagger}_{j} \,
\hat{B}_{m} .. \hat{B}_{1} | \phi_R \rangle }  
{\langle \phi_L \, |  \,\,  \hat{B}_{N_{\tau}} \dots  \hat{B}_1 \,\, | \phi_R \rangle}\,. 
\end{split}
\end{equation}
The central result is provided by the following:
\begin{equation}
\label{assaad_theorem}
\mathcal{G}^p(n, m) = \mathcal{G}^p(n, n-1) \mathcal{G}^p(n-1, n-2) \dots \mathcal{G}^p(m+1, m)
\end{equation}
and:
\begin{equation}
\label{jump_properties}
\mathcal{G}^p(l, l-1) = \mathcal{G}^p(l, l) \mathcal{B}_l
\end{equation}
where $\mathcal{B}_l \equiv \mathcal{B}(\bf{\tilde{x}}(l))$ is the matrix of Eq.~\eqref{b_matrix}.
These identies are straightforward to show using the commutators between the propagator and the 
creation or annihilation operators \cite{Fahy,Brenda}. 
Similar relations can be easily obtained for the holes:
\begin{equation}
\label{assaad_theorem2}
\mathcal{G}^h(n, m) = \mathcal{G}^h(m+1, m) \mathcal{G}^h(m+2, m+1) \dots \mathcal{G}^h(n, n-1)
\end{equation}
and
\begin{equation}
\label{jump_properties2}
\mathcal{G}^h(l, l-1) = \left(\mathcal{B}_l \right)^{-1}\mathcal{G}^h(l, l)\,.
\end{equation}
\COMMENTED{
The identities \eqref{assaad_theorem} and \eqref{jump_properties} (respectively \eqref{assaad_theorem2} and \eqref{jump_properties2}) provide a stable algorithm to compute the particles (respectively holes) dynamical Green 
function, in imaginary time.
} 
The composite properties above
 allow one to build dynamical Green functions from shorter segments.
 One can break the original imaginary time interval into segments of arbitrary length,
say $(n,n-k)$, which turns the matrix in Eq.~(\ref{jump_properties}) 
into the product of the $k$ matrices $\mathcal{B}_l$ inside
the segment. 
Multiplying the Green functions, which are physical properties,
instead of composing the $\mathcal{B}_l$ matrices, 
provides more accurate and numerically stable
results for the dynamical correlations even for large values of the imaginary time.
The length of the segments can be chosen for the best balance between numerical stability
(requiring short segments) and computational time (less for longer segments).
The complexity of this approach 
scales with the lattice size as 
$\mathcal{N}_s^3$, since it requires multiplications of matrices of size $\mathcal{N}_s \times \mathcal{N}_s$.

\begin{figure}[ptb]
\begin{center}
\vspace*{-0.8in}
\includegraphics[width=7.0cm, angle=270]{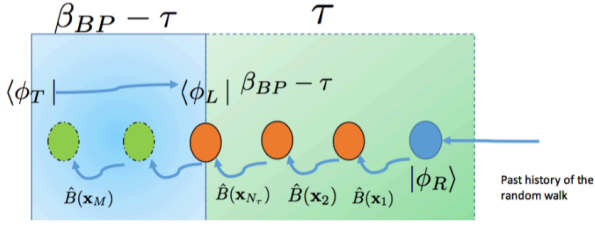}
\vspace*{-0.5in}
\caption{(color online) The calculation of imaginary-time correlation functions in
an open-ended branching random walk: a sketch of the implementation.}
\label{fig:sketch}
\end{center}
\end{figure}

Although we have discussed the formalism in the 
path-integral AFQMC framework, it can be easily modified 
for a constrained
path approach of open-ended imaginary time projection, which becomes necessary
when the sign problem is present.
This is sketched in 
Fig.~\ref{fig:sketch}.
When back-propagation 
begins, one records the initial population and their weights, which play the role of $\{|\phi_R\rangle\}$.  As the random walk proceeds (imposing the constraint along the path as needed), one keeps track of the 
sampled auxiliary-fields 
and their path history,
which gives the path ${\bf X}=({\bf x}(1), \cdots, {\bf x}(M))$ for each descendent walker at time $\beta_{\rm BP}\equiv M\delta\tau$ 
beyond when BP begins. We must choose $\beta_{\rm BP}$ such that $\beta_{\rm BP}-\tau$ is sufficiently large
to project out the ground state from $\langle \phi_T|$. That is, the segment of the path from 
$l\equiv M-N_\tau$ to $M$ is used to produce  $\langle \phi_L|$, using Eq.~\eqref{phi_L}.
The remainder, from ${\bf x}(1)$ to  ${\bf x}(N_\tau)$, play the same 
role as 
in Eq.~\eqref{dynamical estimator}. Since the entire auxiliary-field configurations (or, equivalently, 
segments of the propagator matrices) have been recorded along the BP path, we can proceed in the 
same way as described above to obtain the dynamical Green's functions. As in BP computation of 
equal-time Green's functions, the weight to be used in the Monte Carlo estimator should be that of 
the 
descendent walker at ``future time'' $\beta_{\rm BP}$. \cite{Wirawan-boson}

Once the imaginary-time Green's functions are computed, 
many correlation functions can be obtained. The charge gap can be computed as we discuss below.
Further, the Green's functions are related to 
the spectral function: 
\begin{widetext}
\begin{equation}
\label{spectral_function}
A(Q,\omega) = 
\begin{cases}
\sum_{n}  | \langle  \Psi^{\mathcal{N}_p + 1}_n | \hat{c}^{\dagger}_{Q,\sigma}  |  \Psi_0 \rangle  |^2 \delta\left( \omega - \left(E_{n}^{\mathcal{N}_p + 1} - E^{\mathcal{N}_p}_{0}\right)\right), \quad \omega > \mu \\
\sum_{n}  | \langle  \Psi^{\mathcal{N}_p - 1}_n  | \hat{c}^{}_{Q,\sigma}  |  \Psi_0 \rangle  |^2 \delta\left( \omega + \left(E_{n}^{\mathcal{N}_p - 1} - E^{\mathcal{N}_p}_{0}\right)\right), \quad \omega < \mu
\end{cases}
\end{equation}
\end{widetext}
Here $\Psi^{\mathcal{N}_p \pm 1}_n$ are the eigenstates of the hamiltonian operator
with $\mathcal{N}_p \pm 1$ particles corresponding to the energies  $E_{n}^{\mathcal{N}_p \pm 1}$, while
$\mu$ is the chemical potential.
The spectral function is proportional to the imaginary
part of the time-ordered Green function. 
It can be experimentally measured from
photoemission and inverse photoemission spectroscopy. 
The spectral function provides 
insight into the nature of
the single particle spectral weight for a 
correlated 
system, and is a central object in many-body theory.
From the imaginary-time correlation functions, a Fredholm integral equation
has to be solved to determine the spectral function; in practice:
\begin{equation}
\int_{\mu}^{+\infty} \, e^{-\tau \omega} A(Q,\omega) d\omega = 
  \frac{ \langle \Psi_0  | \, \hat{c}^{}_{Q} \, e^{-\tau  (\hat{H} - E_0) }   \hat{c}^{\dagger}_{Q}  \, | \Psi_0 \rangle }
{\langle \Psi_0 \, | \, \Psi_0 \rangle}
\label{eq:spectral-G1}
\end{equation}
and:
\begin{equation}
\int_{-\infty}^{\mu} \, e^{-\tau \omega} A(Q,\omega) d\omega = 
  \frac{ \langle \Psi_0  | \, \hat{c}^{\dagger}_{Q} \, e^{\tau  (\hat{H} - E_0) }   \hat{c}^{}_{Q}  \, | \Psi_0 \rangle }
{\langle \Psi_0 \, | \, \Psi_0 \rangle}\,.
\label{eq:spectral-G2}
\end{equation}

\begin{figure}[ptb]
\begin{center}
\includegraphics[width=6.0cm, angle=270]{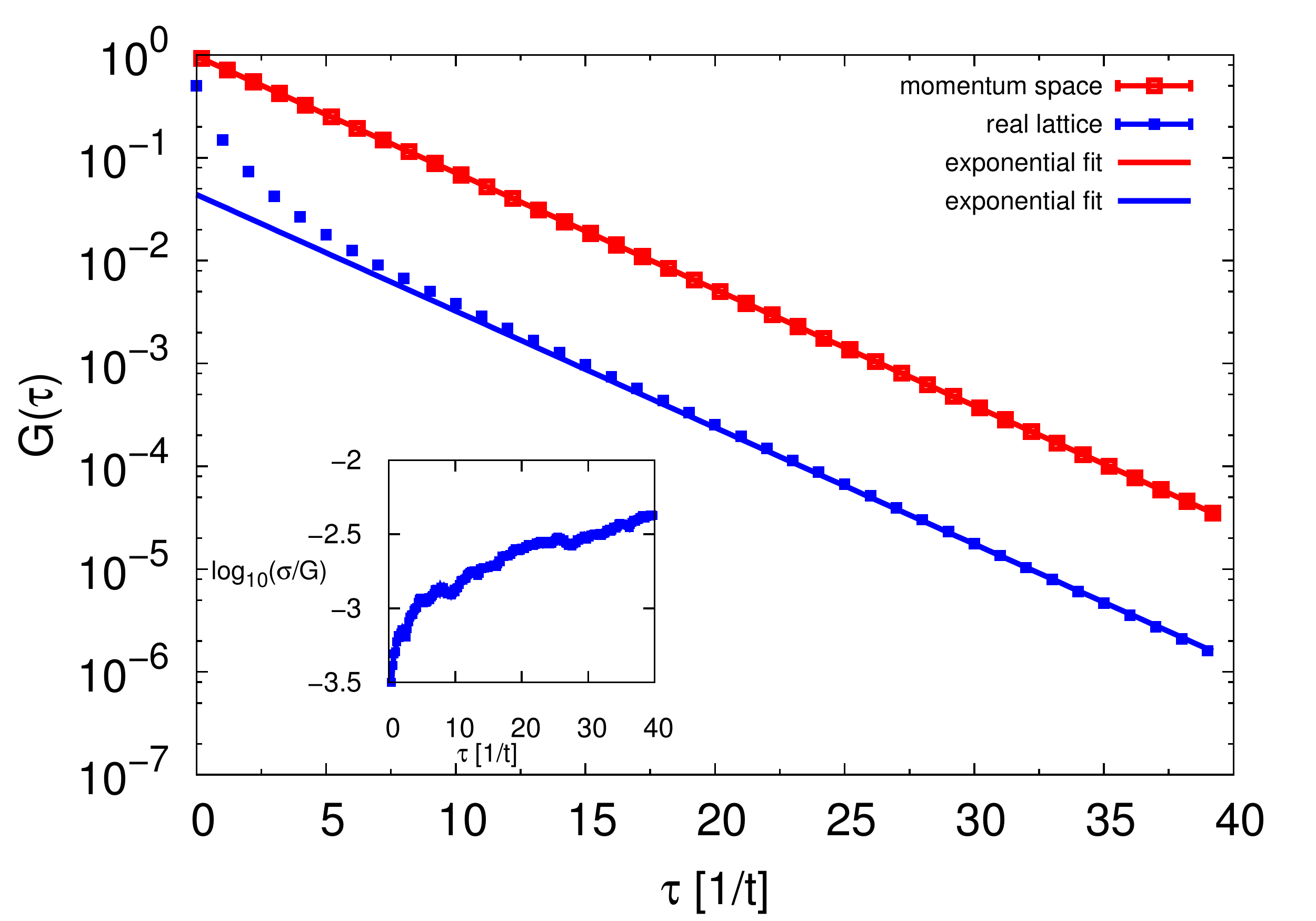}
\caption{(color online) Dynamical Green functions  in real- and momentum-space, and the dependence
of statistical errors on imaginary-time.
$G(R,\tau)$ was computed at $R=0$, and 
$G(Q,\tau)$
at a $Q$ close to
the Fermi surface. 
The system was a  $6 \times 6$ lattice at half-filling with
$U/t = 0.5$.
In the main figure, statistical errors are much smaller than symbol size.
The straight lines are exponential
fits to the large imaginary time region (note semi-log scale). The inset shows the dependence
of the relative error bar on the imaginary time.}
\label{fig:betasign}
\end{center}
\end{figure}

The charge gap $\Delta$ is typically 
inferred from the behavior of the zero distance \emph{real-space} Green function for large imaginary time.
In general, 
\begin{equation}
\Delta = \frac{1}{2}\left(\varepsilon_p + \varepsilon_h\right)
\label{gap:gen_def}
\end{equation}
with:
\begin{equation}
\varepsilon_{p,h} = -\lim_{\tau \to +\infty}\frac{\log\left(G^{p,h}(R=0,\tau)\right)}{\tau} = E(\mathcal{N}_p \pm 1) - E(\mathcal{N}_p)\,,
\end{equation}
where $E(\mathcal{N}_p)$ is the ground state energy for $N$ particles,
while $E(\mathcal{N}_p\pm 1)$ correspond to the lowest energy
eigenstates of the $\mathcal{N}_p \pm 1$ systems having non-zero 
overlap with 
the state obtained by adding/removing a 
particle in any momentum state to the $\mathcal{N}_p$-particle
ground state.
%
At half filling, particle-hole symmetry \cite{hirsch} allows us to simplify the above
definition:
\begin{equation}
\label{gap_config}
\Delta = -\lim_{\tau \to +\infty}\frac{\log\left(G^{p}(R=0,\tau)\right)}{\tau}  - \mu
\end{equation}
Since the hamiltonian defined in Eq.~\eqref{ham} is particle-hole symmetric, the chemical potential  $\mu$ 
is zero at half filling.

We could also compute the gap in \emph{momentum space}:
\begin{equation}
\label{my_idea}
\Delta= \min_{Q}\left(-\lim_{\tau t \to + \infty} \frac{\log\left(\tilde{G}^{p}(Q,\tau)\right)}{\tau}\right)\,. 
\end{equation}
Eq.~\eqref{my_idea} provides an intuitive physical meaning of the charge gap: $\Delta$ is related
to the minimum energy among the unoccupied states, which can be
probed via inverse photoemission
spectroscopy (see, for example,
Ref.~\onlinecite{arpes} and references therein), using a collimated beam of electrons directed
at the sample. 
(At half filling, particle-hole symmetry makes this energy
coincide with the minimum energy needed to extracting a photoelectron
in a direct spectroscopy experiment when the sample is illuminated
via electromagnetic radiation.)

Computing the dynamical Green function in momentum
space is more convenient than in real space, at least for smaller values of $U/t$. 
In this regime $G^{p}(R=0,\tau)$ contains a linear combination of exponentials
while $\tilde{G}^{p}(Q,\tau)$ has a simpler
structure which can be handled more easily 
via analytic continuation
methodologies.
It is straightforward to see this in the limiting case of $U=0$, when $\tilde{G}^{p}(Q,\tau)$ is 
a single exponential:
\begin{equation}
\tilde{G}_{0}^{p}(Q,\tau) =  \overline{n}_{0}(Q)  \, e^{-\tau \varepsilon(Q)}\,,
\end{equation}
where $ \overline{n}_{0}(Q)= 1 - n_{0}(Q)$, with $n_{0}(Q)$ being the Fermi distribution.
In contrast, 
\begin{equation}
\label{idgas:expansion}
G_{0}^{p}(R=0,\tau) =  \frac{1}{\mathcal{N}_s} \sum_{Q} \, \overline{n}_{0}(Q) \, e^{-\tau \varepsilon(Q)}\,,
\end{equation}
in which the minimum gap ($\varepsilon(Q_F) = 0$ in this case, $Q_F$ belonging to the
Fermi surface) provides the charge gap at 
sufficiently large $\tau$ but many other exponentials can persist for significant $\tau$ values, especially as the system size grows.

In fact we could construct a linear combination to target in order to optimize convergence and 
statistical accuracy in the computation of the gap: 
\begin{equation}
\tilde{G}_B^{p}(\tau) \propto \sum_{Q \in B} \, \tilde{G}^{p}(Q,\tau)\,.
\end{equation}
One simple definition for the momentum domain in the summation could be 
$ B=\left\{Q:\,  | \overline{n}(Q) -  \overline{n}_0(Q) | > {\rm const} \right\}$,
where $\overline{n}(Q) = 1 - {n}(Q)$ is defined with respect to the momentum distribution
of the interacting system,  ${n}(Q)$.
The value of ${\rm const}$ can be tuned.

Even more generally, one could use any single
particle orbital $|\mu\rangle$ to create a quasi particle
excitation. The real and momentum space shown above
are simply two special cases. In the formalism presented above, suitable linear combinations of the dynamical
Green functions would be required.
This possibility can be particularly useful in the new approach we propose in 
Sec.~\ref{sec:new-method}, where any 
single-particle orbital (for example a natural orbital obtained from the many-body calculation), 
can be propagated along with the ground-state random walker
with little additional cost.

In Fig.~\ref{fig:betasign} we show an example of calculation 
of imaginary-time Green functions 
at half filling with $U/t=0.5$, in both real space and momentum space.
It is evident that the two reach the
same slope in logarithmic scale at large imaginary time but that $\tilde{G}^{p}(Q,\tau)$
has a much simpler structure, allowing us to accurately calculate
the slope without the need of reaching very large imaginary
times.
This is important since
 the relative statistical
uncertainty increases exponentially, as shown in the inset. 

\begin{figure}[ptb]
\begin{center}
\includegraphics[width=6.0cm, angle=270]{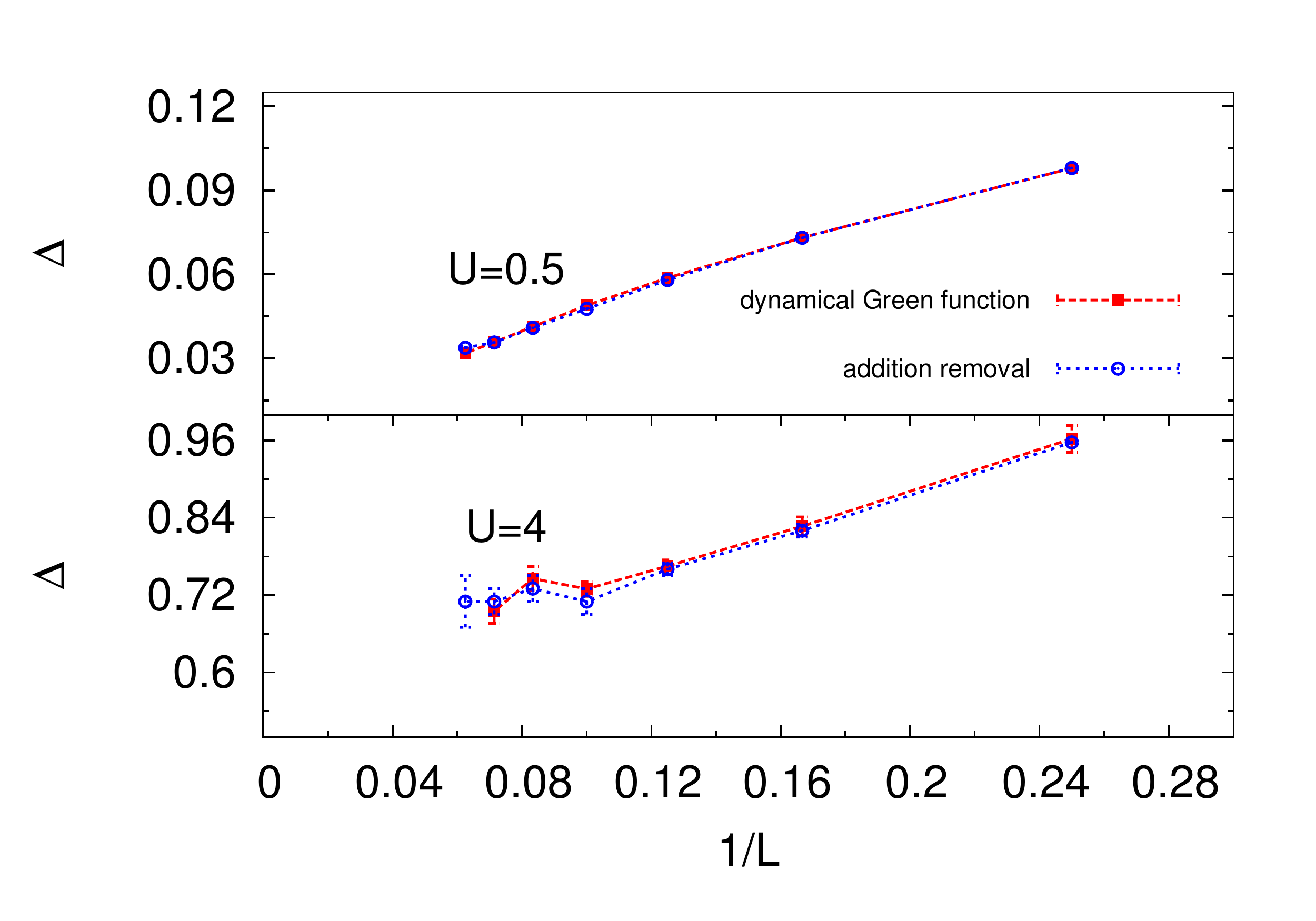}
\caption{(color online) Charge gap measured from dynamical Green function (filled squares) and
from addition/removal (open circles), as a function of the inverse linear size of the system, $1/L$.
The two panels are for two different interaction strengths:  $U=0.5$ (upper) and $U=4$ (lower). 
}
\label{fig:gap_0.5_pbc_ar}
\end{center}
\end{figure}

The
charge gap can also be estimated with an addition/removal technique, 
calculating directly $\varepsilon_p$ and  $\varepsilon_h$ in Eq.~\eqref{gap:gen_def}.
This approach has the advantage that
it does not require the evaluation of dynamical correlation functions.
It also has several disadvantages. 
With three separate calculations (or two, if at
half-filling) of $\mathcal{N}_p$ and
$\mathcal{N}_p\pm 1$ particles, 
it involves the difference between extensive quantities, which can give rise to large statistical uncertainties for large systems. 
Second, while not present in Hubbard-like models, the addition/removal of an electron in 
a supercell in real solids tends to create a significant additional finite-size effects which requires larger
supercells or better correction schemes in the many-body calculation. 
Moreover, in the Hubbard model at half-filling the systems with $\mathcal{N}_p\pm1$ particles both have a
sign problem, while the (half-filled) $\mathcal{N}_p$ system does not \cite{hirsch}. 

We have performed  addition/removal  calculations of the gap to help 
check the robustness of the imaginary-time Green's function approach. 
These calculations were performed with the constrained path Monte Carlo (CPMC) method
\cite{cpprl1995,AFQMC-lecture-notes-2013}. 
We used trial wave functions obtained from the generalized  Hartree-Fock (GHF), which 
were found to improve the results near half-filling  \cite{hub-bench,mingpu-to-be-published}.
For the $(\mathcal{N}_p-1)$ calculation, we use the same GHF trial wave 
function as for half-filling, simply omitting the extra orbital in the minority-spin sector,
while the next virtual GHF orbital is used in the $(\mathcal{N}_p+1)$ case.
This was found, by comparison with 
exact diagonalization results, to give exact results on the gap to within statistical errors. 
 
In Fig.~\ref{fig:gap_0.5_pbc_ar}  the gaps computed from imaginary-time Green's function and from addition/removal are compared for a variety of systems ranging from $4\times 4$ to $16\times 16$
periodic supercells. 
Excellent agreement is seen between the two approaches. 
It is also evident that the gap converges rather slowly with supercell size. Especially at smaller values
of $U$, 
very large lattices are needed 
and a fit in $1/L$ is difficult and can be unreliable. We discuss how to obtain more robust 
estimates at the thermodynamic limit in the next section.



\section{Gaps at the thermodynamic limit}


\subsection{Approaches to reduce the finite-size effects}

\begin{figure}[ptb]
\begin{center}
\includegraphics[width=6.0cm, angle=270]{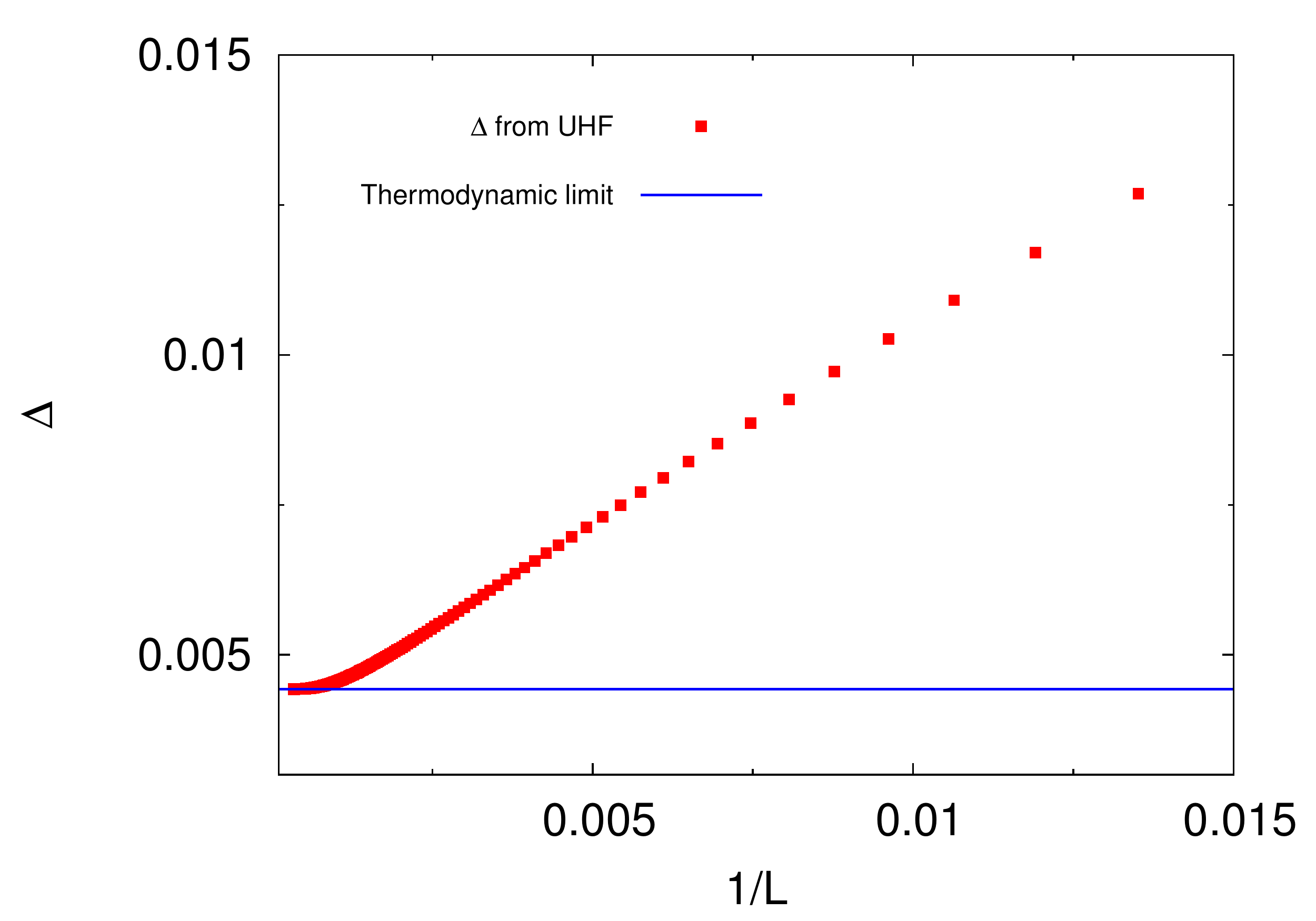}
\caption{(color online) Charge gap measured from  UHF calculations at $U=0.5$, as a function of $1/L$.
The straight line indicates the UHF gap value at the thermodynamic limit, 0.0044272.
}
\label{fig:gap_0.5_pbc_uhf}
\end{center}
\end{figure}

To access bulk properties it is crucial to be able to extrapolate the results to the thermodynamic limit.
We perform
simulations up to lattices containing $24\times 24$ sites, i.e. $576$ electrons,
more than twice 
the largest lattice for which dynamical calculations had been performed before.
Nevertheless, as shown in Fig.~\ref{fig:gap_0.5_pbc_ar}, it is still challenging to determine the gap reliably, especially for small values
when it is necessary to resolve the existence of a gap.
In order to 
shed light into the behavior for $1/L \to 0$, we performed
unrestricted Hartree Fock (UHF) calculations systematically as a function of lattice size. 
The gaps are obtained by the orbital energies in a Koopman's theorem type of approach for each 
supercell size. The results are shown in Fig.~\ref{fig:gap_0.5_pbc_uhf}. Note that most of the 
$L$ values in the data are beyond reach of many-body calculations in supercells 
with present day computing power. 
The quantitative (or even qualitative) accuracy of UHF aside, the 
results 
illustrate the strong finite-size effects that must be overcome in order to reach the   
 thermodynamic limit accurately.

\COMMENTED{
not trivial to infer the correct dependence on the size of the system.
In the upper panel of  Fig.~\ref{fig:gap_0.5_pbc_ar} we plot the estimated gap
at $U/t = 0.5$  for lattices 
with $4\times 4$, $6\times 6$, ... up to $16\times 16$ sites, in periodic boundary conditions
as a function of $1/L$. We show the estimation from dynamical correlation function together the addition-removal estimation. 

A close inspection to Fig.~\ref{fig:gap_0.5_pbc_ar} reveals a downward curvature of the data, and a polynomial fit is difficult. In order to shed light into the behavior for $1/L \to 0$, we performed
Unrestricted Hartree Fock (UHF) calculations for very large systems.
Even if we cannot be sure that UHF provides the correct scaling
of the charge gap, the behavior of the data in Fig.~\ref{fig:gap_0.5_pbc_uhf} strongly suggests that a robust estimation of the thermodynamic limit is difficult
if we can only reach sizes such that $1/L \gtrsim 1/24$.
}

Our first step to reduce finite-size effects is to use twisted boundary conditions
with quasi-random sequences of twist parameters \cite{mingpu-to-be-published}.
Formally, the introduction of a twist parameter $\theta=(\theta_x,\theta_y)$, $0 \leq \theta_{x,y} < 1$ means that the algebra of creation and destruction operators satisfy the
new boundary conditions:
\begin{equation}
\hat{c}^{\dagger}_{i + L \hat{x}} = e^{i 2\pi \theta_x} \, \hat{c}^{\dagger}_{i}, \quad
\hat{c}^{\dagger}_{i + L \hat{y}} = e^{i 2\pi \theta_y} \, \hat{c}^{\dagger}_{i}
\end{equation}
for all sites $i$; $\hat{x}$ and $\hat{y}$ denote the unit vectors in $x$ and $y$ directions.
This implies that the wave vectors are $Q_{x,y}= \frac{2\pi}{L}(n_{x,y} + \theta_{x,y})$ or, 
equivalently, that we can replace the original dispersion relation $\varepsilon(Q)$
with $\varepsilon(Q+\frac{2\pi}{L}\theta)$.

\begin{figure}[ptb]
\begin{center}
\includegraphics[width=6.cm, angle=270]{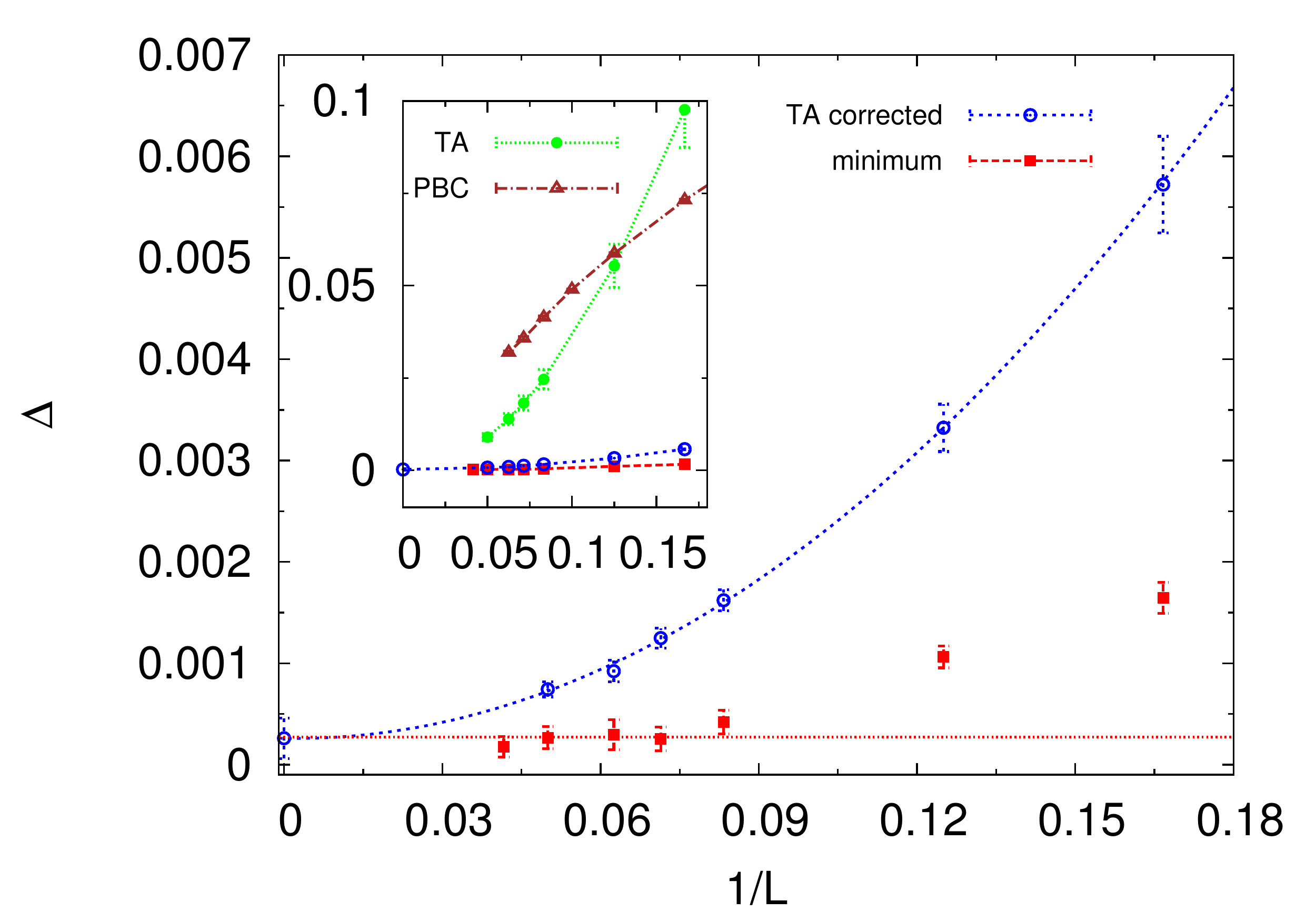}
\caption{(color online) Reduction of the finite-size effects and convergence to the thermodynamic limit in 
computing the
charge gap. The charge gap at $U/t = 0.5$ measured from dynamical Green function are shown
as a function of $1/L$, from 
a twist-averaging (TA) procedure together with one-body correction (empty circles)
and taking the minimum among the corrected gaps  (filled squares).
The dotted line is a quadratic fit to the twist-averaged data.
The straight line is the estimation of the thermodynamic limit, obtained
using the minimum gap estimator performing a linear fit.
The inset shows the same data, together with 
the results from TA prior to the one-body correction are also shown  (filled circles) together with 
those from (PBC) (filled triangles), which contain large finite-size effects. 
}
\label{fig:gap_0.5_tbc}
\end{center}
\end{figure}

It is known \cite{PhysRevE.64.016702} that performing averages of physical quantities, like
the ground state energy, with respect to the
twist parameter strongly reduces finite-size effects. The discretization
of the Fermi ``sphere'' due to the finite size is 
smeared out by
the presence of the twist parameters. 
A straightforward application of twist averaging in our calculations 
yields the results shown in the inset of Fig.~\ref{fig:gap_0.5_tbc}
for $U/t = 0.5$.
The error bars are estimated as a combination of the uncertainty
from the analytic continuation for a given twist and the one coming from the twist averaging (obtained
with a jackknife estimator).
From the plot it is evident that the role of the boundary conditions
is important. 

\begin{figure}[ptb]
\begin{center}
\vspace*{-0.5in}
\includegraphics[width=6.5cm, angle=270]{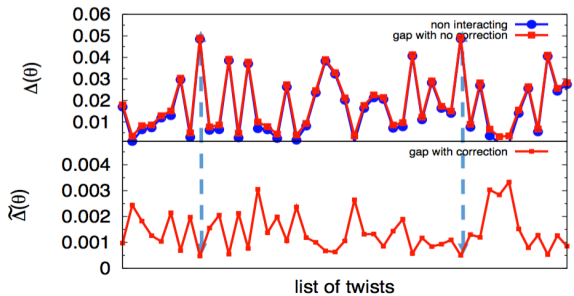}
\vspace*{-0.5in}
\caption{(color online) 
Finite-size effects in computing the gap, the use of twist boundary conditions, and special 
twist values. The top panel shows  
the non-interacting gap, and the exact many-body gap 
as a function of the twist parameters.
The bottom panel shows the corrected gaps, and identifies the minimum. 
The system is a $14 \times 14$ lattice at $U/t=0.5$.
}
\label{fig:gap_of_twist}
\end{center}
\end{figure}

We can further improve convergence to the thermodynamic limit by removing the one-body 
finite-size effect \cite{PhysRevB.59.1917,PhysRevLett.100.126404}. 
We can correct 
the values of the computed gap for a
given $\theta$ by the \emph{finite-size} non-interacting gap, 
to eliminate or reduce the 
effects arising purely from the shift of the Fermi sphere and the shell structure. 
The upper panel of  Fig.~\ref{fig:gap_of_twist} illustrates this effect, where a strong correlation is evident
between the true many-body gap $ \Delta(\theta) $ and the non-interacting gap $\Delta^{0}(\theta)$.
The simple correction
$\tilde \Delta(\theta) = \Delta(\theta) +\left(\Delta^{0}-\Delta^{0}(\theta) \right)$,  
where $\Delta^{0}=0$ is the non-interacting gap at the thermodynamic limit, 
gives one order of magnitude reduction in the fluctuations of the gap values.
This is expected at small $U$. We find that, 
although the dependence on the twist parameter
is weaker at, say, $U=4$, the correlation is also present at larger $U$.
The gap result as a function of system size after the one-body finite-size correction 
is shown 
in Fig.~\ref{fig:gap_0.5_tbc}. 
We see that the twist-averaged result (averaging $\tilde \Delta(\theta)$ over the twist $\theta$,
in this case nearly $50$ quasi random twist parameters)
 has much smaller error bars, because of the reduction in the fluctuations
between twist parameters mentioned above, and
displays better convergence and
a quadratic scaling
in $1/L$. 
\COMMENTED{
actually is almost linear in $1/L^2$, and the gap converges to 
a non-zero thermodynamic limit. This is noticeable since the energy scale
is tiny, but we have enough resolution to find a non-zero result, confirming
the picture that, at half-filling, the ground state of the repulsive Hubbard model is insulating even for small values of $U/t$.
 }
 
 We propose a third step for accelerating convergence, involving a different way to
use the twisted boundary conditions than the more standard procedure above.  
After one-body finite-size correction, 
we seek the minimum gap among all the twist angles
\begin{equation}
\label{my_idea_correction}
\Delta_{\rm min} = \min_{\theta}\left( \tilde\Delta(\theta)  \right)\,. 
\end{equation}
This allows one to better sample for the minimum in Eq.~\eqref{my_idea}.
\COMMENTED{
we argue that the introduction of twist parameters
allows for an improvement in looking for the minimum
 We will use the notation
$\Delta(\theta)$ for the value of the gap for 
a given twist parameter. We expect that the twist
parameters allow us to better sample the function:
\begin{equation}
Q  \to -\lim_{\tau t \to + \infty} \frac{\log\left(\tilde{G}^{p}(Q,\tau)\right)}{\tau}
\end{equation}
which, in the thermodynamic limit, becomes a function of a continuous
variable, while, for a finite system, it is defined on a discrete lattice.
We expect that, introducing a set of twist parameters, we can have
}
We stress that the difference between Eq.~\eqref{my_idea_correction}, averaging
over $\tilde \Delta(\theta)$, and even $\Delta(\theta)$ itself vanishes in the thermodynamic limit. 
However, taking the minimum among the post-correction twist results enable
access to estimations of the values of Eq.~\eqref{my_idea} for a much larger
set of points, which improves the estimation of the minimum.
As we see in Fig.~\ref{fig:gap_0.5_tbc}, this procedure leads to a further improvement over
the TA results, with the computed gap becoming essentially  flat for $L \gtrsim 12$.
Most of our results in the next section are obtained with this procedure of finding the minimum 
corrected gap. The statistical uncertainties on the minimum corrected gap are estimated simply as the QMC errors; we have checked that
this estimation is reliable by performing, in some situations, several independent
calculations and computing the variance of $\Delta_{\rm min}$.

\COMMENTED{
Again, we take into account that, for the non interacting
system,  a calculation of \eqref{my_idea} for $\theta \neq (0,0)$
would give, for a finite system, a non zero value. We denote
$\Delta^{0}(\theta)$ such value, which is a pure finite
size effect, which we would like to correct for. To this aim, we
evaluate:

\begin{equation}
\label{my_idea_correction}
\Delta = \min_{\theta}\left( \Delta(\theta) - \Delta^{0}(\theta) \right) 
\end{equation}
where the label $0$ denotes, as before, the non-interacting Green function. 
The exact meaning of \eqref{my_idea_correction} is as follows: we perform simulations for several twist parameters; for all $\theta$, we estimate the charge gap, subtract the non interacting gap, and, finally, look for the minimum with respect to $\theta$. 

We stress that the difference between \eqref{my_idea_correction} 
and \eqref{gap_config} goes to zero in the thermodynamic limit. 

We have performed measurements of the charge gap for nearly $50$ quasi random twist parameters.
}

We have examined the location of $\Delta_{\rm min}$. 
In the bottom panel of Fig.~\ref{fig:gap_of_twist} we have marked the twist angle which yielded 
the minimum. We find that 
the corrected gap $\tilde \Delta(\theta)$ 
reaches its minimum when the non-interacting gap is maximum, that is
around $\theta_{\rm min}=(0,1/2)$ or symmetry-related points.
This observation holds for all the cases we have studied, 
including a variety of lattice sizes $L$ at multiple values of $U$ ($0.5$, $1$, and $4$).
We rationalize the observation 
as follows: when a particle is added at half-filling, 
the system accommodates it by creating a spin wave. The value
$\theta_{\rm min}=(0,1/2)$ allows the maximum wavelength for this excitation,
thus allowing the minimum energy.
In the next section, we apply this special $\theta_{\rm min}$ to obtain the gap value in many other $U$ values 
between $0.5$ and $4$.
Of course 
the particular value of $\theta_{\rm min}$ will depend on the system, but any insight 
towards identifying its value or narrowing its range will help reduce the computational cost.

\COMMENTED{ 
is certainly
related to the choice of the half-filled Hubbard model as
a model system. For calculations on different systems, 
a preliminary investigation of the dependence of the 
gap on the twist angle will be needed; anyway, we find
very interesting the fact that properly tuning the boundary
conditions, the finite-size effects are dramatically reduced
}

\subsection{Results on gaps and the spectral function for the Hubbard model at half-filling}

In this section, we present our results  of the charge gap in the repulsive Hubbard model at half-filling,
systematically as a function of the interaction, as well as the spectral function at $U=4t$. 
These calculations are similar to prior efforts 
that exist to study such quantities, using both 
 the Lanczos method for 
 small lattices \cite{scalapino, oppenheimer} and QMC for larger lattices
 \cite{assaad_prl,hirsch,assaad_recent}. 
 Our calculations reach larger system sizes and apply the approaches discussed above to 
 systematically reach the thermodynamic limit. 
 
 \COMMENTED{
 
 In the case of the Hubbard model in two dimensions at zero temperature, 
the spectral function has been calculated
for small lattices with the Lanczos method\cite{scalapino, oppenheimer}.
Assaad \cite{assaad_prl} et al. studied the spectral function at zero temperature for lattices 
up to $16\times 16$ using AFQMC techniques, in order to explore the insulator-metal
transition.

In the context of finite temperature studies, calculations for the Hubbard model have been performed via QMC methods
by Hirsch \cite{hirsch}; more recently,  Assaad et. al. performed determinantal
quantum Monte Carlo calculations at finite temperature, discussing extrapolation
to the thermodynamic limit and estimating the pseudo-gap.

}

\begin{figure}[ptb]
\begin{center}
\includegraphics[width=6.0cm, angle=270]{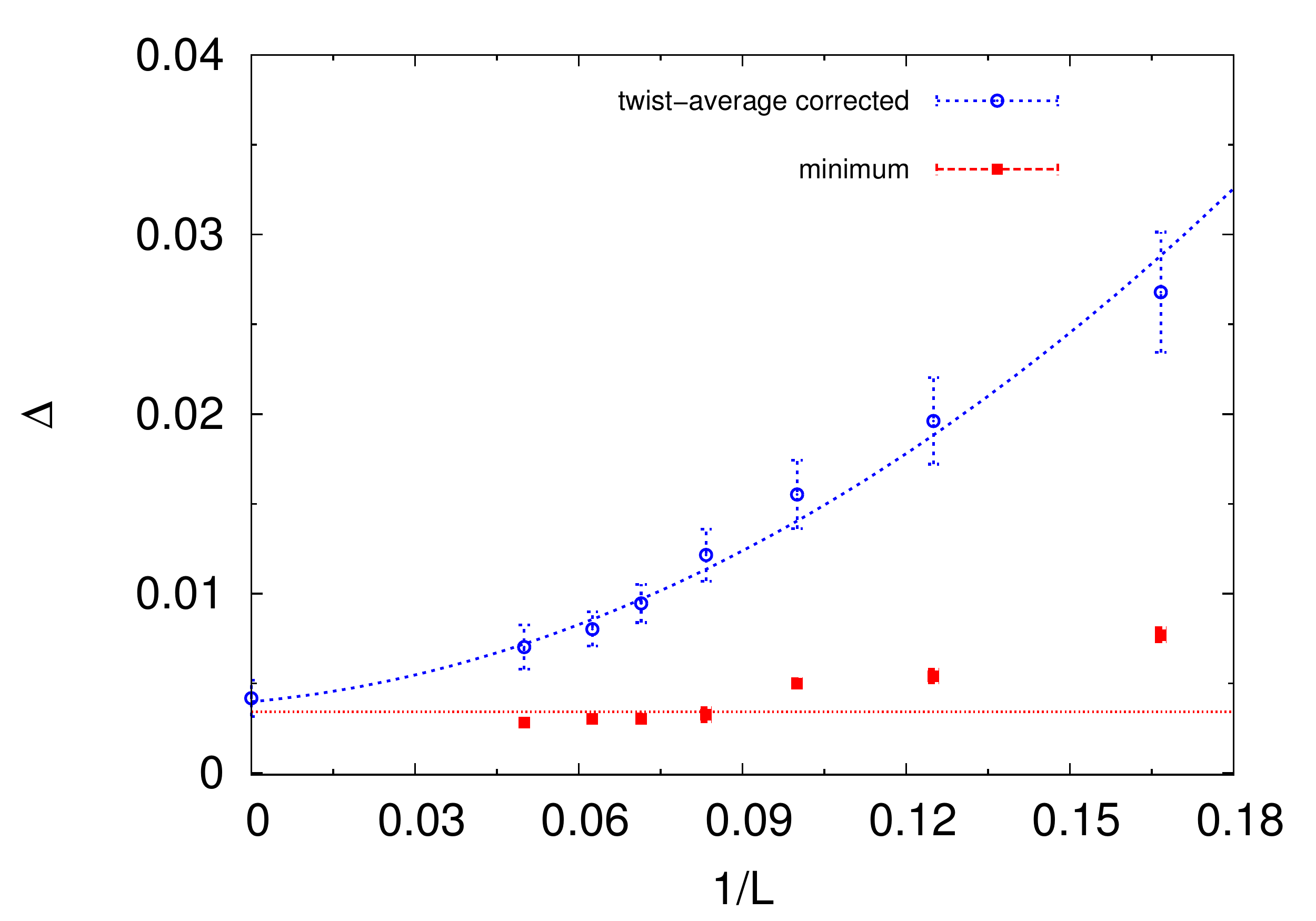}
\caption{(color online) Charge gap at $U/t = 1$ vs.~the inverse (linear) system size. 
The gas are measured from dynamical Green function
with twist-averaging and one-body correction (empty circles)
and with the minimum $\Delta_{\rm min}$ (filled squares).
The dotted line is a quadratic fit to the twist-averaged data.
The straight line is the estimation of the thermodynamic limit, obtained
using the minimum gap estimator.}
\label{fig:gap_1_tbc}
\end{center}
\end{figure}

\begin{figure}[ptb]
\begin{center}
\includegraphics[width=6.0cm, angle=270]{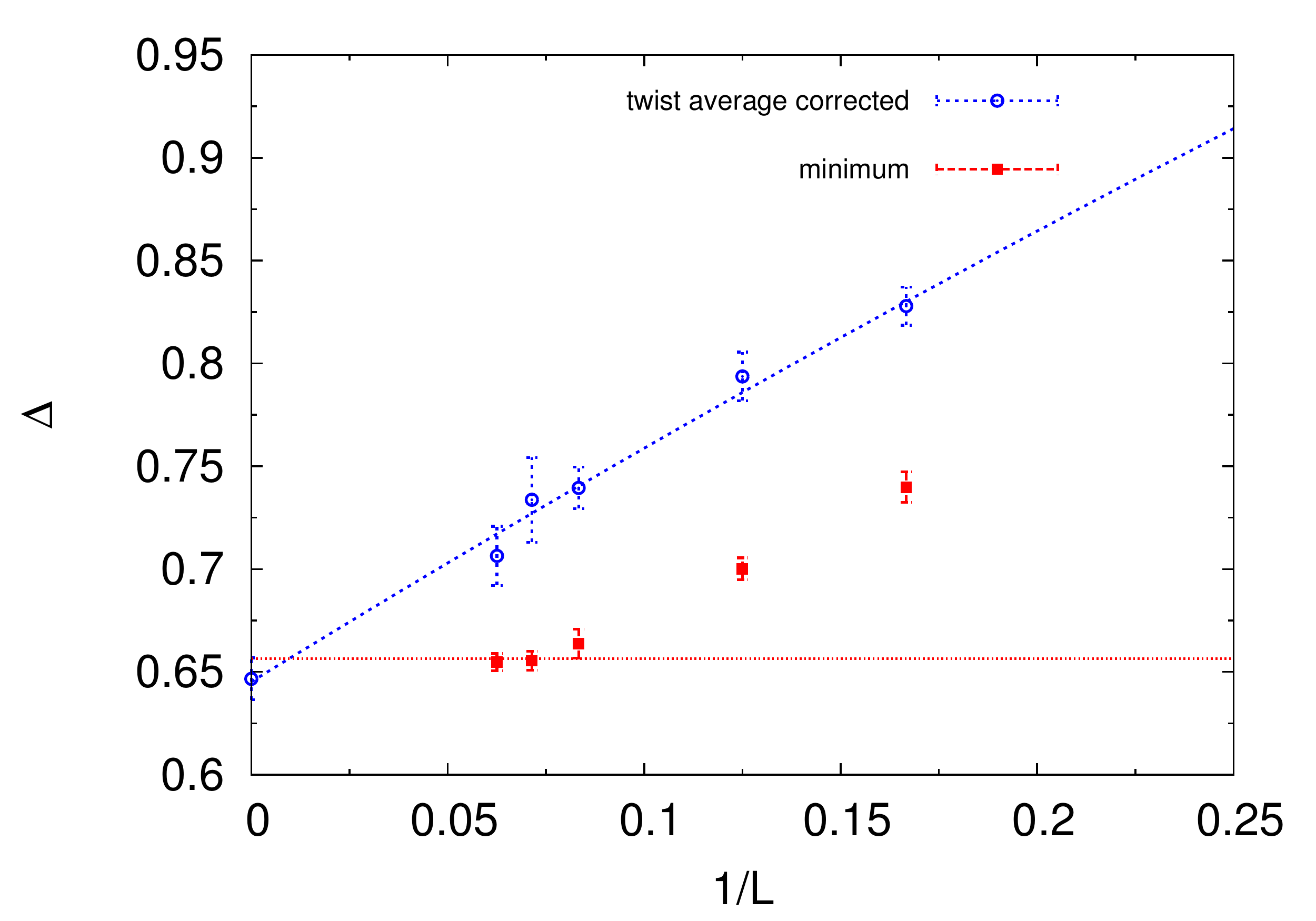}
\caption{(color online) Charge gap at $U/t = 4$. Symbols and setup are the same as in Fig.~\ref{fig:gap_1_tbc}.
\COMMENTED{
measured from dynamical Green function
with a twist-averaging procedure together with one-body correction (empty circles)
and minimum among the several gaps  (filled squares)
as a function of $1/L$, $L$ being the linear size of the system.
The straight line is the estimation of the thermodynamic limit, obtained
using the minimum gap estimator.}
}
\label{fig:gap_4_tbc}
\end{center}
\end{figure}

\COMMENTED{

In this paper, since our primary interest is to quantitatively discuss
the extrapolation to the thermodynamic limit, instead of
studying the whole spectral function, we focus on the calculation of the charge gap, 
related to the minimum distance between the
particles ($\omega > \mu$) and holes ($\omega < \mu$) 
peaks in the spectral function \eqref{spectral_function}.
We will show the result of the calculation of the full spectral
function at the end of this section, in Fig.~\ref{fig:ako_4_tbc}.
A calculation of the spectral function in the Hubbard model in two dimensions,
at the presence of doping, will be the topic of a future paper.

We perform the calculations when the sign problem is not present, 
estimating the charge gap at half-filling, i.e. $N_{\uparrow}=N_{\downarrow}=\frac{1}{2}\mathcal{N}_s$, for $U>0$. 
}

In Fig.~\ref{fig:gap_0.5_tbc} we show the final results obtained for $U=0.5t$.
Lattice sizes up to $24 \times 24$ were studied using multiple quasi-random twist angles. 
\COMMENTED{
We are aware
of calculations of dynamical properties from QMC simulations for systems
of up to $16 \times 16$ sites  \cite{assaad_recent}, so that with current calculations we reach more than
twice the size that had been studied before. 
We show both the estimation obtained using \eqref{my_idea_correction} and
the result of the twist averaging procedure. 
The estimator
\eqref{my_idea_correction} becomes nicely flat, for $L \gtrsim 12$.
}
We find a charge gap of 
 $\Delta = 0.00027(4)$  at the thermodynamic limit. 
This
very small value is clearly impossible to determine using conventional  calculations with periodic boundary conditions (PBC) or even twist-averaged boundary conditions (TABC).
That the gap value is small but non-zero is significant, confirming 
 that the ground state of the Hubbard
model is insulating 
at small finite $U$.

In Fig.~\ref{fig:gap_1_tbc} we show the results of the same calculation for $U/t=1$.
Qualitatively the behavior is the same as for $U=0.5$. The statistical uncertainties
are larger, since $U=0.5$ is so small that the dynamical Green functions are 
very similar to the ones of the non-interacting system, which makes the finite-size
correction in Eq.~\eqref{my_idea_correction} especially effective in reducing the statistical
uncertainty. Our estimation of the charge gap at  $U/t=1$ is $\Delta = 0.0034(4)$.


We present results for $U/t=4$ in Fig.~\ref{fig:gap_4_tbc}.
In this case, the twist averaged results displays a nearly linear dependence
on $1/L$, consistent with
results from a previous QMC study \cite{assaad_prl}. 
As in the other cases, our estimator from Eq.~\eqref{my_idea_correction} becomes flat also in this case.
Our estimate of the charge gap at $U/t=4$ is $\Delta = 0.656(2)$.

\COMMENTED{
An interesting question is whether there is some regular pattern
in the dependence of the estimator
\eqref{my_idea_correction} on the twist parameter.
In Fig.~\ref{fig:gap_of_twist} we show, for a finite system at $U/t=0.5$, the
non-interacting gap, the interacting gap and the difference between
the two, i.e. the estimator \eqref{my_idea_correction}, as a function
of $\theta$. It is evident that the interacting gap follows quite closely
the behavior of the non-interacting gap.
This is expected at small $U$, and we found that
this behavior is present also at larger $U$, although,
for $U=4$, the dependence on the twist parameter
is weaker.
{\color{red}On the other hand, the corrected gap $\Delta(\theta) - \Delta^{0}(\theta)$,
reaches its minimum when the non-interacting gap is maximum, that is
around $\theta=(0,\frac{1}{2})$ or symmetry-related points.
This can be understood as follows: when we add a particle, 
the system accomodates if creating a spin wave; this value
of $\theta$ allows the maximum wavelength for this excitation,
thus allowing the minimum energy.
We observe that the particular value of $\theta$ is certainly
related to the choice of the half-filled Hubbard model as
a model system. For calculations on different systems, 
a preliminary investigation of the dependence of the 
gap on the twist angle will be needed; anyway, we find
very interesting the fact that properly tuning the boundary
conditions, the finite-size effects are dramatically reduced.}
}

\begin{figure}[ptb]
\begin{center}
\includegraphics[width=6cm, angle = 270]{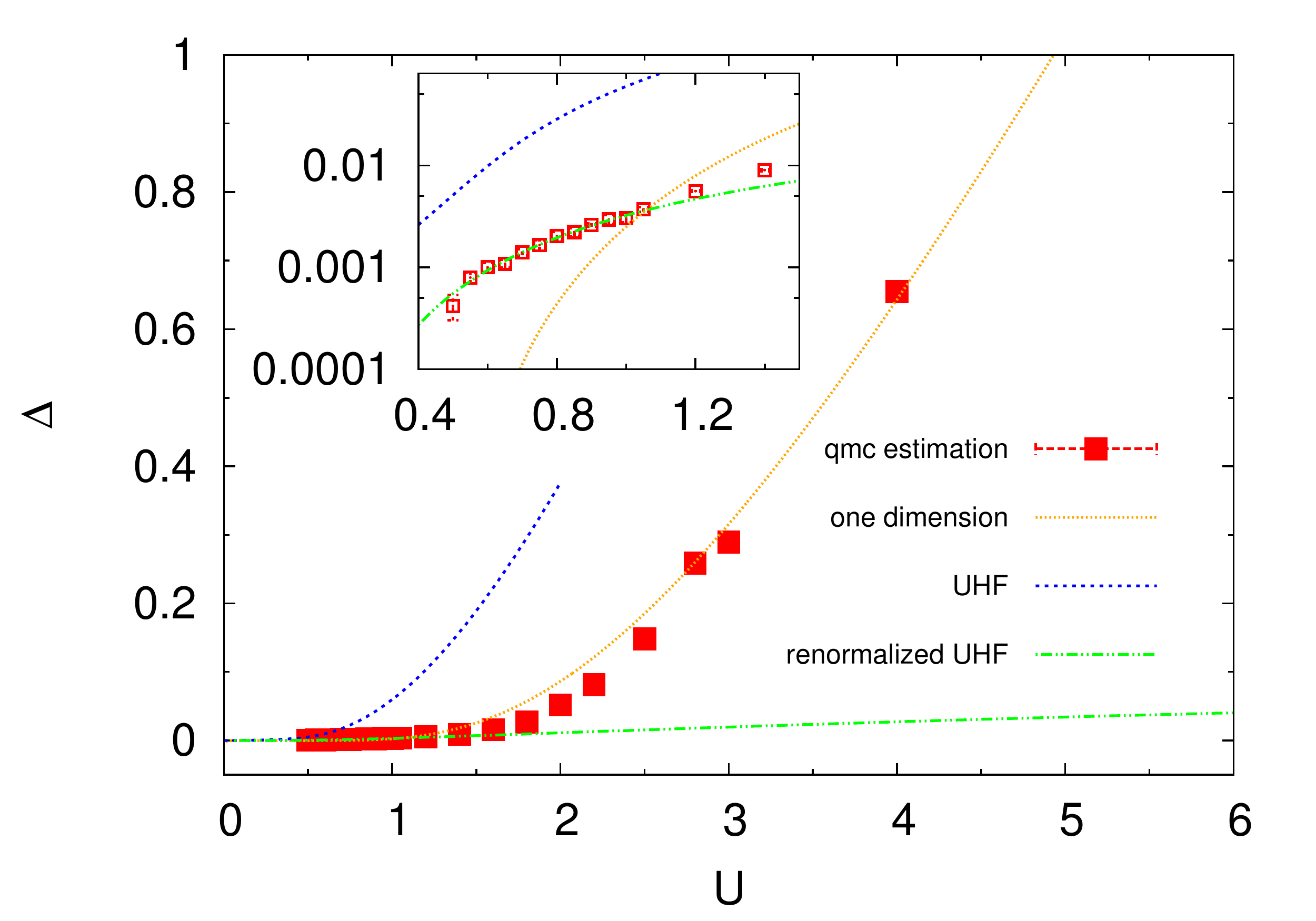}
\caption{(color online) Gap at half-filling as a function of the interaction strength. 
Symbols are obtained from AFQMC calculations. Statistical error bars are shown but are smaller 
than symbol size. 
The (green) dashed line corresponds to a fit of the QMC data with a mean-field form allowing 
renormalized parameters. The (blue) dotted line is the actual mean-field result from unrestricted Hartree-Fock.
The (orange) line 
at large $U$ is the Bethe ansatz prediction for one-dimension. 
The inset shows a zoom of the main graph at small $U$.
}
\label{fig:gap_of_U}
\end{center}
\end{figure}

Next, we map out a detailed $\Delta$ vs.~$U$ curve by studying a variety of 
interaction strengths.
Having established the $\theta_{\rm min}$ value from the systematic searches at the $U$ values 
studied above, we now
 use the special twist value $\theta_{\rm min}$ for each additional system, computing the gap at 
 a sufficiently large $L$. We have verified in a few systems by calculations at multiple $L$'s that 
 convergence to the thermodynamic limit has been reached.  
%
%
The results 
are shown in Fig.~\ref{fig:gap_of_U}.
We find that, at small $U$, the gap behaves
as $\Delta(U) = \alpha \exp\left( - \beta \sqrt{\frac{t}{U}}\right)$, as predicted
by UHF \cite{hirsch}, with renormalized parameters $\alpha$ and $\beta$. 
We find $\alpha = 0.23(4)$ and $\beta = 4.3(2)$, in contrast
with the UHF predictions $\alpha = 32$ and $\beta = 2\pi$.
(The actual UHF results are seen to, not surprisingly, severely over-estimate the gap.)
On the other hand, at large $U/t$, the gap appears to follow the same behavior
predicted in \emph{one-dimension} from the Bethe ansatz \cite{lieb-wu}:
\begin{equation}
\label{bethe}
\Delta_{1D}(U) = \frac{U}{2} - 2 + 4 \int_{0}^{+\infty} d\omega \frac{J_1(\omega)} 
{ \omega (1 + \exp (\omega U/2) )}\,,
\end{equation}
where $J_1(\omega)$ is a Bessel function.

\COMMENTED{
In Fig.~\ref{fig:gap_of_U} we show the results for many values
of the interaction $U/t$. We find that, at small $U$, the gap behaves
as $\Delta(U) = \alpha \exp\left( - \beta \sqrt{\frac{t}{U}}\right)$, as predicted
by UHF \cite{hirsch}, with renormalized parameters $\alpha$ and $\beta$. 
On the other hand, at large $U/t$, the gap appears to follow the same behavior
that can be exactly predicted in one-dimension using the Bethe ansatz \cite{lieb-wu}:
\begin{equation}
\label{bethe}
\Delta_{1D}(U) = \frac{U}{2} - 2 + 4 \int_{0}^{+\infty} d\omega \frac{J_1(\omega)} 
{ \omega (1 + \exp (\omega U/2) )}
\end{equation}
where $J_1(\omega)$ is a Bessel function.
}

\begin{figure}[ptb]
\begin{center}
\includegraphics[width=8.5cm, angle=0]{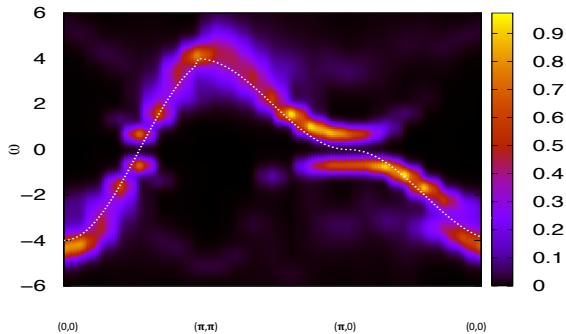}
\caption{(color online) 
Color plot of the spectral function $A(Q,\omega)$ as a function of
momentum $Q$ (horizontal axis) along the principal directions
in the Brillouin zone  and frequency $\omega$ (vertical axis). 
The spectral function has been obtained by performing analytic continuation
of the calculated imaginary time Green functions in
momentum space. The system was a $16\times16$ lattice
at $U/t=4$. The dotted line is the non-interacting dispersion relation. 
}
\label{fig:ako-4-tbc}
\end{center}
\end{figure}

Finally, in Fig.~\ref{fig:ako-4-tbc} we show a calculation of the full spectral function
$A(Q,\omega)$ defined in Eq.~\eqref{spectral_function}. 
The horizontal axis shows $Q$  along a path in the Brillouin zone, indicated by the end points of the
straight-line segments.
The spectral function was obtained from analytic continuation of the imaginary-time 
Green functions, as discussed in Eqs.~\eqref{eq:spectral-G1} and \eqref{eq:spectral-G2}.
The analytic continuation has been performed using the Genetic Inversion
via Falsification of Theories (GIFT) method \cite{gift}. 
As mentioned, the spectral function 
can be measured
experimentally via photoemission experiments, and 
provides a map of
the single particle states of the system.
From the plot, where the
non-interacting spectral function $A^{0}(Q,\omega) = \delta(\omega - \varepsilon(Q))$ is also shown,
the effect of the interactions is evident, with the opening of a gap at the
Fermi surface, as well as the broadening and renormalization of the dispersion
relation.


\section{New Method}
\label{sec:new-method}

In this section we propose a new method
which will enable faster computations of the Green's functions in a larger number of situations than the method of 
Sec.~\ref{ssec:method-dynamical}. In particular, the new method allows
us to
calculate single matrix elements of the Green's function
with a number of operations scaling linearly with lattice sites (or basis size).
In systems such as dilute Fermi gas and 
\emph{ab initio} calculation of realistic systems \cite{phaseless2003,AFQMC-lecture-notes-2013},
the lattice or basis size is significantly larger than the number of particles,
  $\mathcal{N}_s \gg \mathcal{N}_p$, so that it is advantageous to be able to calculate the Green functions
(and certain correlation functions) with computational cost of $\mathcal{O}(\mathcal{N}_s\,\mathcal{N}_p^2) $
versus $\mathcal{O}(\mathcal{N}^3_s)$.
Even if the calculation of the full Green's function 
matrix would still require the latter, 
generally we are interested in in a subset of them, not all the elements.
Moreover, as we will show below, the
method we are proposing allows to extend the calculation in
a straightforward way to two-body dynamical correlations without 
affecting the complexity.

\subsection{Particle excitations}
\label{ssec:new-method-particle}
We consider the estimator in Eq.~\eqref{dynamical estimator}
in a more general sense:
\begin{equation}
\label{dynamical estimator_new_gen}
g_{\nu,\mu}({\bf{X}},\tau) =\frac{ \langle \phi_L  | \,  \hat{c}^{}_{\nu} \,\hat{B}_{N_{\tau}} \dots \hat{B}_1  \,\,\hat{c}^{\dagger}_{\mu}  | \phi_R \rangle }
{\langle \phi_L \, |  \,\hat{B}_{N_{\tau}} \dots \hat{B}_1  \,\,| \, \phi_R \rangle }\,,
\end{equation}
where $|\nu\rangle$ and $|\mu\rangle$ are single-particle
orbitals, which can be either position eigenstates $|i\rangle$, $|j \rangle$, 
momentum eigenstates $|Q\rangle$ 
or even more general states such as natural orbitals. 

\COMMENTED{
In order to keep the notations
simple will consider the case of a single operator $\hat{B}$, the
generalization being a straightforward iteration.
}

Let us start from the $\mathcal{N}_p$-particles Slater determinant: 
\begin{equation}
| \phi_R \rangle = \hat{c}^{\dagger}_{u_1} \dots \hat{c}^{\dagger}_{u_{\mathcal{N}_p}} \, |0\rangle\,,
\label{eq:psiR}
\end{equation}
where $\hat{c}^{\dagger}_{u} = \sum_i \langle i | u \rangle \, \hat{c}^{\dagger}_{i}$ creates a particle 
in the orbital $|u\rangle$. 
It is convenient to assume that the orbitals $|u_1\rangle, \dots, |u_{\mathcal{N}_p}\rangle$ form an
orthonormal set, which in practice is realized by, for example, a modified 
Gram-Schmidt (GS) procedure.
The creation operator $\hat{c}^{\dagger}_{\mu}$
adds to $| \phi_R \rangle$ one particle in the orbital $|\mu \rangle$,
giving rise to a new $(\mathcal{N}_p+1)$-particle Slater determinant:
\begin{equation}
\label{npu}
| \phi^{\mathcal{N}_p+1}_R \rangle = \hat{c}^{\dagger}_{\mu} \, | \phi_R \rangle\,.
\end{equation}
%
Written in orthonormal form:
\begin{equation}
\label{npu2}
| \phi^{\mathcal{N}_p+1}_R \rangle = \hat{c}^{\dagger}_{\tilde{\mu}}  \hat{c}^{\dagger}_{u_1} \dots \hat{c}^{\dagger}_{u_{\mathcal{N}_p}} \, |0\rangle\, \mathcal{D}_{\mathcal{N}_p+1}\,,
\end{equation}
where $|\tilde{\mu}\rangle$ is the orbital  $|\mu\rangle$ after projecting out the linear dependence on 
$\{|u\rangle\}$:
\begin{equation}
\label{primastab}
|\tilde{\mu}\rangle = \frac{\mu - \sum_{\alpha=1}^{\mathcal{N}_p} \langle u_\alpha \, | \mu \, \rangle \, | u_\alpha \rangle}
{|| \mu - \sum_{\alpha=1}^{\mathcal{N}_p} \langle u_\alpha \, | \mu \, \rangle \, | u_\alpha \rangle ||}, 
\end{equation}
with $\mathcal{D}_{\mathcal{N}_p+1} = 
\langle \tilde{\mu} \, | \mu \rangle$.
Now, if a propagator $\hat{B}$ is applied,  we have:
\begin{equation}
\hat{B} \, | \, \phi_R \rangle =   \hat{c}^{\dagger}_{\hat{B}u_1} \dots \hat{c}^{\dagger}_{\hat{B} u_{\mathcal{N}_p}} \, |0\rangle\,,
\end{equation}
and
\begin{equation}
\label{bnpu2}
\hat{B}| \phi^{\mathcal{N}_p+1}_R \rangle = \hat{c}^{\dagger}_{\hat{B}\tilde{\mu}}  \hat{c}^{\dagger}_{\hat{B}u_1} \dots \hat{c}^{\dagger}_{\hat{B}u_{\mathcal{N}_p}} \, |0\rangle\, \,\mathcal{D}_{\mathcal{N}_p+1}\,.
\end{equation}
That is, each orbital is propagated by the  one-particle propagator $ \hat{B}$,
\COMMENTED{
both the original $\mathcal{N}_p$-particle determinant and the new, $(\mathcal{N}_p+1)$-particle one evolve in the same way, with the one-particle propagator $ \hat{B}$ operating on each orbital.
\begin{equation}
| u_{\alpha} \rangle \leadsto \hat{B} | u_{\alpha} \rangle, \quad 
| \tilde{\mu} \rangle \leadsto \hat{B} | \tilde{\mu} \rangle\,,
\end{equation}
}
so that both the $\mathcal{N}_p$- and the $(\mathcal{N}_p+1)$-particle Slater determinants remain in form as
Slater determinants.
%
In orthonormal form:
\begin{equation}
\hat{B} \, | \, \phi_R \rangle = |\, \phi'_R \,\rangle \, \mathcal{D}
\label{eq:bphirgs1}
\end{equation}
with
\begin{equation}
\label{bphirgs}
 |\, \phi'_R \,\rangle =  \hat{c}^{\dagger}_{u'_1} \dots \hat{c}^{\dagger}_{u'_{\mathcal{N}_p}} \, |0\rangle\,,
\end{equation}
where the orbitals $|u'_1\rangle, \dots, |u'_{\mathcal{N}_p}\rangle$ form an
orthonormal set and $\mathcal{D}$ is the factor arising from the GS
decomposition, 
and
\begin{equation}
\label{bphirnpugs}
\hat{B}| \phi^{\mathcal{N}_p+1}_R \rangle = | \phi^{\mathcal{N}_p+1\,\prime}_R \rangle \,\, \mathcal{D} \,\, \mathcal{D}'_{\mathcal{N}_p+1} \, \mathcal{D}_{\mathcal{N}_p+1}
\end{equation}
with:
\begin{equation}
 | \phi^{\mathcal{N}_p+1\, \prime}_R  \rangle \, = \hat{c}^{\dagger}_{\tilde{\mu}'} | \phi^{\prime}_R \rangle =   \, \hat{c}^{\dagger}_{\tilde{\mu}'}  \hat{c}^{\dagger}_{u'_1} \dots \hat{c}^{\dagger}_{u'_N} \, |0\rangle
\end{equation}
where, as before,   
$|\tilde{\mu}'\rangle$ is the orbital  $|\hat{B}\tilde{\mu}\rangle$ after projecting out the linear dependence on 
$\{|u'\rangle\}$, and 
$ \mathcal{D}'_{\mathcal{N}_p+1} = 
\langle \tilde{\mu}' \, | \hat{B}\tilde{\mu} \rangle$. 
\COMMENTED{
$|u'_1\rangle, \dots, |u'_N\rangle$  and  $\mathcal{D}$ are the same as in
\eqref{bphirgs}, $\mathcal{D}_{\mathcal{N}_p+1}$ is defined in \eqref{primastab}, while,
as before:

\begin{equation}
\label{primastab2}
|\tilde{\mu}'\rangle = \frac{\hat{B}\tilde{\mu} - \sum_{\alpha=1}^{\mathcal{N}_p} \langle u'_\alpha \, | \hat{B}\tilde{\mu} \, \rangle \, | u'_\alpha \rangle}
{|| \hat{B}\tilde{\mu} - \sum_{\alpha=1}^{\mathcal{N}_p} \langle u'_\alpha \, | \hat{B}\tilde{\mu} \, \rangle \, | u'_\alpha \rangle ||}, \quad \mathcal{D}'_{\mathcal{N}_p+1} = 
\langle \tilde{\mu}' \, | \hat{B}\tilde{\mu} \rangle 
\end{equation}
}

This procedure can be iterated for the product of $\hat B$  in Eq.~\eqref{dynamical estimator_new_gen}, propagating the single-particle orbitals and accumulating the weights 
$\mathcal{D}'$. 
As in regular AFQMC for static observables, 
the GS reorthonormalization need not be applied at every iteration, only with a frequency sufficient to 
ensure numerical stability  \cite{cpprl1995}.
 The evaluation of the Green's function element requires the calculation of:
\begin{equation}
g_{\nu\mu} ({\bf{X}},\tau) = \mathcal{D}'_{\mathcal{N}_p+1}\mathcal{D}_{\mathcal{N}_p+1} \, 
\frac{ \langle \phi_L  | \,  \hat{c}^{}_{\nu} \, \phi^{\mathcal{N}_p+1\,\prime}_R \rangle  }
{\langle \phi_L \, | \phi'_R \rangle  } \,.
\end{equation}
Applying $ \hat{c}^{}_{\nu}$ to the left, we can evaluate the numerator as the overlap of two 
 $(\mathcal{N}_p+1)$-particle Slater determinants.
Equivalently:
\begin{equation}
\label{eq:g-dynamic-particle-2}
g_{\nu\mu}({\bf{X}},\tau) = 
\mathcal{D}'_{\mathcal{N}_p+1}\mathcal{D}_{\mathcal{N}_p+1} \, 
\frac{ \langle \phi_L  | \,  \hat{c}^{}_{\nu} \,  \hat{c}^{\dagger}_{\tilde{\mu}'} \, | \,  \phi^{\prime}_R \rangle  }
{\langle \phi_L \, | \phi'_R \rangle  }\,.
\end{equation}
In explicit matrix representation,
if we denote:
\begin{equation}
\Phi'_R = \left(
\begin{array}{ccc}
 \langle 1 | u'_1 \rangle &  \dots  &  \langle 1 | u'_{\mathcal{N}_p} \rangle \\
 \langle 2 | u'_1 \rangle &  \dots & \langle 2 | u'_{\mathcal{N}_p} \rangle  \\
 \langle 3 | u'_1 \rangle &  \dots &  \langle 3 | u'_{\mathcal{N}_p}\rangle \\
  \dots &  \dots  & \dots   \\
    \dots  &   \dots &   \dots  \\
   \langle  \mathcal{N}_s | u'_1 \rangle &   \dots &  \langle  \mathcal{N}_s | u'_{\mathcal{N}_p} \rangle \\
\end{array}
\right)
\label{eq:phipR-mat}
\end{equation}
and
\COMMENTED{
\begin{equation}
\langle \phi_L | = \langle 0 | \,\,  \hat{c}^{}_{v_{\mathcal{N}_p}} \dots  \hat{c}^{}_{v_{\mathcal{N}_p}}  
\end{equation}
}
\begin{equation}
\Phi_L = \left(
\begin{array}{ccc}
 \langle 1 | v_1 \rangle &  \dots  &  \langle 1 | v_{\mathcal{N}_p} \rangle \\
 \langle 2 | v_1 \rangle &  \dots & \langle 2 | v_{\mathcal{N}_p} \rangle  \\
 \langle 3 | v_1 \rangle &  \dots &  \langle 3 | v_{\mathcal{N}_p} \rangle \\
  \dots &  \dots  & \dots   \\
    \dots  &   \dots &   \dots  \\
   \langle  \mathcal{N}_s | v_1 \rangle &   \dots &  \langle  \mathcal{N}_s |v_{\mathcal{N}_p} \rangle \\
\end{array}
\right)
\label{eq:phiL-mat}
\end{equation}
for $\langle \phi_L | = \langle 0 | \,\,  \hat{c}^{}_{v_{\mathcal{N}_p}} \dots  \hat{c}^{}_{v_{\mathcal{N}_p}} $,
simple algebraic manipulations allow us to
obtain the ratio on the right Eq.~\eqref{eq:g-dynamic-particle-2}, the matrix element, as
\COMMENTED{
\begin{equation}
\frac{ \langle \phi_L  | \,  \hat{c}^{}_{\nu} \,  \hat{c}^{\dagger}_{\tilde{\mu}'} \, | \,  \phi^{\prime}_R \rangle  }
{\langle \phi_L \, | \phi'_R \rangle  }
\end{equation}
in the form:
}
\begin{equation} 
\label{new_estimator_part}
\left( \langle \nu \, | \tilde{\mu}'\rangle
- \sum_{\alpha,\beta=1}^{\mathcal{N}_p}   \langle \nu \, | u'_{\alpha}\rangle \, \langle u'_{\beta} \, | \tilde{\mu}'\rangle
\left( \bf{\left(\Phi_{L}\right)^{\dagger} \, \Phi'_{R} }\right)^{-1}_{\alpha,\beta}  \right)\,,
\end{equation}
which can be evaluated with $\mathcal{O}(\mathcal{N}_s\,\mathcal{N}_p^2)$ operations for
a given pair of $|\tilde{\mu}'\rangle$ and $|\nu\rangle$.


\begin{figure}[ptb]
\begin{center}
\includegraphics[width=6cm, angle = 270]{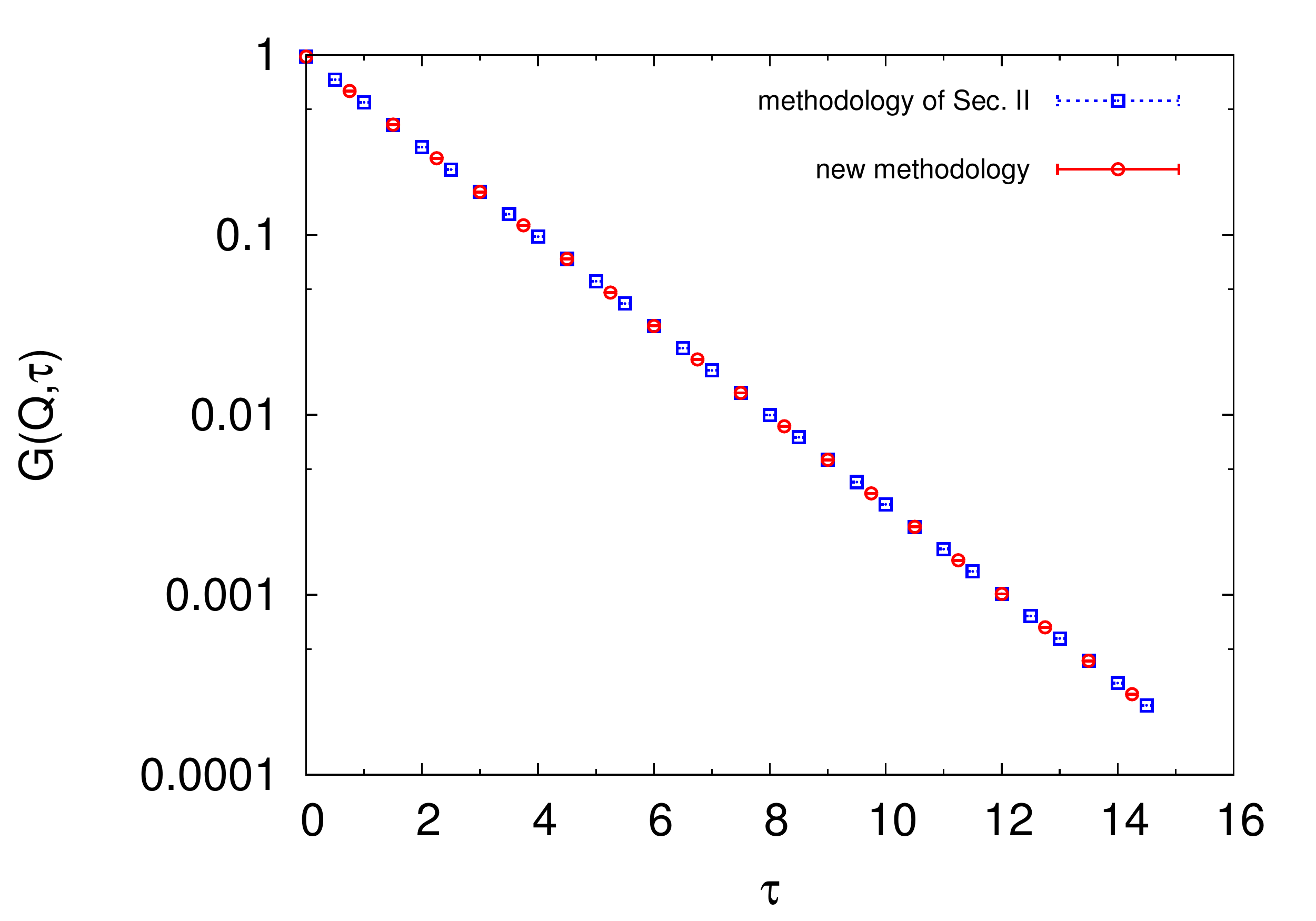}
\caption{(color online) Particle Green function $\tilde{G}(Q,\tau)$ for a large lattice in the repulsive Hubbard model: comparison between  
 the new methodology and that
described  in Sec.~\ref{ssec:method-dynamical}.
A $12 \times 12$ lattice with $72$ spin-$\uparrow$
and $72$ spin-$\downarrow$ particles is studied 
at $U/t = 1$. }
\label{fig:benchmark2}
\end{center}
\end{figure}

In Fig.~\ref{fig:benchmark2} we show a comparison
between the particle Green's function computed using the method discussed in Sec.~\ref{ssec:method-dynamical} and the
one computed with the present method. 
The perfect agreement shows the robustness of both approaches. 
%
As mentioned, 
the 
savings
from the present method 
occur when a small number of  Green's function elements are targeted. 
When the whole Green's function is needed, the approach becomes more computationally expensive.
Also, the advantage relies on  $\mathcal{N}_p \ll \mathcal{N}_s$, so that
at half-filling as in Fig.~\ref{fig:benchmark2}, for  
example, there is  no advantage over the previous method. 
On the other hand,  in the dilute limit the reduction 
 in computational cost (and potentially in memory requirement as well, since
the propagated determinants are of size  $\mathcal{O}(\mathcal{N}_s\mathcal{N}_p)$
vs.~$\mathcal{O}(\mathcal{N}_s^2)$ for $\mathcal{G}$) can be dramatic,
such as in systems of cold atoms \cite{Hao-2DFG}
and especially with spin-orbit coupling present \cite{PhysRevLett.117.040401} which further increases the basis
size. 

\COMMENTED{
, for a particular Green's function element,
once the couple of sites $(i,j)$ is fixed, 
the method described in Sec.~\ref{ssec:method-dynamical}  requires 
$\mathcal{O}(\mathcal{N}^3_s)$ operations, while the present method provides
the result with $\mathcal{O}(\mathcal{N}_s)$ operations. We observe that
this simplification occurs when the particles number satisfies $\mathcal{N}_p << \mathcal{N}_s$,
since the estimator \eqref{new_estimator_part} contains cicles over the
number of particles. For example the new estimator is not useful at
half-filling, but becomes extremely useful in the dilute limit,
which is notoriously important for the physics of cold atoms \cite{Hao-2DFG, PhysRevLett.117.040401}.

In Fig.~\ref{fig:benchmark2} we show another example of the comparison
between the two methodologies for a much larger system. It appears
clearly that the new methodology does not suffer from numerical instability problems
at large values of the imaginary time.
}

\subsection{Hole excitations}
Suppose now we wish to compute the estimator:
\begin{equation}
\label{dynamical estimator_new_gen_h}
h_{\nu,\mu}({\bf{X}},\tau) =\frac{ \langle \phi_L  | \,  \hat{c}^{\dagger}_{\nu} \,\hat{B}_{N_{\tau}} \dots \hat{B}_1  \,\,\hat{c}^{}_{\mu}  | \phi_R \rangle }
{\langle \phi_L \, |  \,\hat{B}_{N_{\tau}} \dots \hat{B}_1  \,\,| \, \phi_R \rangle }\,.
\end{equation}
\COMMENTED{
Again, to keep notations simple, we consider just one operator
$\hat{B}$, the general case being a simple iteration of the
algorithm we will describe.
Assuming the same {eq:psiR}
as before that the orbitals $|u_1\rangle, \dots, |u_{\mathcal{N}_p}\rangle$
building the Slater determinant $| \phi_R \rangle $  form an
orthonormal set, we let:
}
We again consider $\hat{c}^{}_{\mu}$ acting on $ | \phi_R \rangle$:
\begin{equation}
\label{nmu}
| \phi^{\mathcal{N}_p-1}_R \rangle  =  \hat{c}^{}_{{\mu}} \, | \, \phi_R \rangle 
= \hat{c}^{}_{{\mu}} \,   \hat{c}^{\dagger}_{u_1} \dots \hat{c}^{\dagger}_{u_{\mathcal{N}_p}} \, |0\rangle\,,
\end{equation}
which is no longer a single Slater determinant, but a linear combination of 
${\mathcal{N}_p}$ Slater determinants. However, because these 
determinants all have the same structure, each containing 
$({\mathcal{N}_p}-1)$ orbitals taken from   $ | \phi_R \rangle$,
their propagation by a propagator 
$ \hat{B}$ can be handled together in a convenient way.

We re-write Eq.~\eqref{nmu} as
%
\begin{equation}
\label{nmu2}
| \phi^{\mathcal{N}_p-1}_R \rangle = \hat{c}^{}_{\tilde{\mu}}  \hat{c}^{\dagger}_{u_1} \dots \hat{c}^{\dagger}_{u_{\mathcal{N}_p}} \, |0\rangle\,,
\end{equation}
where $|\tilde{\mu}\rangle$ is the projection of $|\mu\rangle$ onto
the linear space spanned by $|u_1\rangle, \dots, |u_{\mathcal{N}_p}\rangle$:
\begin{equation}
\label{muproj}
|\tilde{\mu}\rangle = \sum_{\alpha=1}^{\mathcal{N}_p} \langle \mu \, | \, u_\alpha \rangle \, | \, u_\alpha \rangle\,.
\end{equation}
\COMMENTED{
The component of $|\mu\rangle$ orthogonal to $|\tilde{\mu}\rangle$ 
has in fact no effect on \eqref{nmu}: keeping only the component $|\tilde{\mu}\rangle$
increases the numerical stability of the algorithm.
}
The propagation of $| \, \phi_R \rangle$ in the denominator in Eq.~\eqref{dynamical estimator_new_gen_h}
is identical to that in the previous section, Sec.~\ref{ssec:new-method-particle}.
\COMMENTED{
\begin{equation}
\hat{B} \, | \, \phi_R \rangle =   \hat{c}^{\dagger}_{\hat{B}u_1} \dots \hat{c}^{\dagger}_{\hat{B} u_{\mathcal{N}_p}} \, |0\rangle
\end{equation}
while, as it follows from simple algebraic manipulations:
}
For the numerator, it follows from simple algebraic manipulations \cite{Brenda}:
\begin{equation}
\label{bnmu2}
\hat{B}| \phi^{\mathcal{N}_p-1}_R \rangle = \hat{c}^{}_{\left(\hat{B}^{-1}\right)^{\dagger}\, |\tilde{\mu}\rangle}  \hat{c}^{\dagger}_{\hat{B}u_1} \dots \hat{c}^{\dagger}_{\hat{B}u_{\mathcal{N}_p}} \, |0\rangle
\end{equation}
The orbital $|\tilde{\mu}\rangle$ evolves under the
action of $\left(\hat{B}^{-1}\right)^{\dagger}$, since it corresponds to 
a destruction operator. 

\COMMENTED{
Again, we apply a Gram-Schmidt decomposition, writing:

\begin{equation}
\hat{B} \, | \, \phi_R \rangle = |\, \phi'_R \,\rangle \,\, \mathcal{D}
\end{equation}
with:

\begin{equation}
\label{bphirgs_bis}
 |\, \phi'_R \,\rangle =  \hat{c}^{\dagger}_{u'_1} \dots \hat{c}^{\dagger}_{u'_{\mathcal{N}_p}} \, |0\rangle
\end{equation}
where he orbitals $|u'_1\rangle, \dots, |u'_{\mathcal{N}_p}\rangle$ form an
orthonormal set and $\mathcal{D}$ is the factor arising from the
decomposition. Contemporarily, we let:
}

After reorthonormalization of in Eqs.~\eqref{eq:bphirgs1} and \eqref{bphirgs}, we can re-write 
\begin{equation}
\label{bphirnmugs}
\hat{B}| \phi^{\mathcal{N}_p-1}_R \rangle = | \phi^{\mathcal{N}_p-1\,\prime}_R \rangle \,\, \mathcal{D} 
\end{equation}
with:
\begin{equation}
 | \phi^{\mathcal{N}_p-1\, \prime}_R  \rangle \, = \hat{c}^{}_{\tilde{\mu}'} | \phi^{\prime}_R \rangle =   \, \hat{c}^{}_{\tilde{\mu}'}  \hat{c}^{\dagger}_{u'_1} \dots \hat{c}^{\dagger}_{u'_{\mathcal{N}_p}} \, |0\rangle\,,
\end{equation}
where $|\tilde{\mu}'\rangle$ is the projection of the orbital $(\hat{B}^{-1})^{\dagger}\, |\tilde{\mu}\rangle$
onto the linear space of $\{ |u'\rangle \}$,
as  in Eq.~\eqref{muproj}.
\COMMENTED{
$|u'_1\rangle, \dots, |u'_{\mathcal{N}_p}\rangle$  and  $\mathcal{D}$ are the same as in
\eqref{bphirgs_bis}. Moreover, as in \eqref{muproj}, we keep only the component:
\begin{equation}
\label{muproj2}
|\tilde{\mu}'\rangle = \sum_{\alpha=1}^{\mathcal{N}_p} \langle \left(\hat{B}^{-1}\right)^{\dagger}\, \tilde{\mu}  \, | \, u'_i \rangle \, | \, u'_i \rangle
\end{equation}
}
The evaluation of the hole Green's function in Eq.~\eqref{dynamical estimator_new_gen_h}  
now becomes:
%
\begin{equation}
h_{\nu\mu} ({\bf{X}},\tau) = 
\frac{ \langle \phi_L  | \,  \hat{c}^{\dagger}_{\nu} \, \phi^{\mathcal{N}_p+1\,\prime}_R \rangle  }
{\langle \phi_L \, | \phi'_R \rangle  } 
=
\frac{ \langle \phi_L  | \,  \hat{c}^{\dagger}_{\nu} \,  \hat{c}^{}_{\tilde{\mu}'} \, | \,  \phi^{\prime}_R \rangle  }
{\langle \phi_L \, | \phi'_R \rangle  }\,.
\end{equation}
\COMMENTED{
or, equivalently:

\begin{equation}
h_{\nu\mu} =  
\frac{ \langle \phi_L  | \,  \hat{c}^{\dagger}_{\nu} \,  \hat{c}^{}_{\tilde{\mu}'} \, | \,  \phi^{\prime}_R \rangle  }
{\langle \phi_L \, | \phi'_R \rangle  }
\end{equation}
If we denote:

\begin{equation}
\Phi'_R = \left(
\begin{array}{ccc}
 \langle 1 | u'_1 \rangle &  \dots  &  \langle 1 | u'_{\mathcal{N}_p} \rangle \\
 \langle 2 | u'_1 \rangle &  \dots & \langle 2 | u'_{\mathcal{N}_p} \rangle  \\
 \langle 3 | u'_1 \rangle &  \dots &  \langle 3 | u'_{\mathcal{N}_p} \rangle \\
  \dots &  \dots  & \dots   \\
    \dots  &   \dots &   \dots  \\
   \langle  \mathcal{N}_s | u'_1 \rangle &   \dots &  \langle  \mathcal{N}_s | u'_{\mathcal{N}_p} \rangle \\
\end{array}
\right)
\end{equation}
and, if:

\begin{equation}
\langle \phi_L | = \langle 0 | \,\,  \hat{c}^{}_{v_{\mathcal{N}_p}} \dots  \hat{c}^{}_{v_1}  
\end{equation}

\begin{equation}
\Phi_L = \left(
\begin{array}{ccc}
 \langle 1 | v_1 \rangle &  \dots  &  \langle 1 | v_{\mathcal{N}_p} \rangle \\
 \langle 2 | v_1 \rangle &  \dots & \langle 2 | v_{\mathcal{N}_p} \rangle  \\
 \langle 3 | v_1 \rangle &  \dots &  \langle 3 | v_{\mathcal{N}_p} \rangle \\
  \dots &  \dots  & \dots   \\
    \dots  &   \dots &   \dots  \\
   \langle  \mathcal{N}_s | v_1 \rangle &   \dots &  \langle  \mathcal{N}_s |v_{\mathcal{N}_p} \rangle \\
\end{array}
\right)
\end{equation}
}
Simple algebraic manipulations 
in the matrix representation of Eqs.~\eqref{eq:phipR-mat} and \eqref{eq:phiL-mat}
allow us to express the matrix element on the right as:
\COMMENTED{
\begin{equation}
\frac{ \langle \phi_L  | \,  \hat{c}^{\dagger}_{\nu} \,  \hat{c}^{}_{\tilde{\mu}'} \, | \,  \phi^{\prime}_R \rangle  }
{\langle \phi_L \, | \phi'_R \rangle  }
\end{equation}
in the form:
}
\begin{equation} 
 \sum_{\alpha,\beta=1}^{\mathcal{N}_p}   \langle  \tilde{\mu}' \, | u'_{\alpha}\rangle \, \langle u'_{\beta} \, | \nu \rangle
\left( \bf{\left(\Phi_{L}\right)^{\dagger} \, \Phi'_{R} }\right)^{-1}_{\alpha,\beta}\,. 
\end{equation}
\COMMENTED{
that provides, for a given couple of orbitals $|\nu\rangle$ and
$|\mu \rangle$, an algorithm to evaluate \eqref{dynamical estimator_new_gen_h}
requiring $\mathcal{O}(\mathcal{N}_s)$ operations.
}

\subsection{Two-body correlation functions}
Suppose we wish to compute a density-density
or a spin-spin correlation function.  In both cases, we have to compute:
\begin{equation}
\label{dynamical estimator_new_gen_tb}
n({\bf{X}},\tau) =\frac{ \langle \phi_L  | \, \hat{n}_{j,\sigma'}  \,\hat{B}_{N_{\tau}} \dots \hat{B}_1  \,\,\hat{n}_{i,\sigma}  | \phi_R \rangle }
{\langle \phi_L \, | \,\hat{B}_{N_{\tau}} \dots \hat{B}_1  \,\,| \, \phi_R \rangle }
\end{equation}
where $\hat{n}_{i,\sigma} = \hat{c}^{\dagger}_{i,\sigma}  \hat{c}^{}_{i,\sigma} $
is the fermion density operator.
In this case, we will write explicitly the spin index.

\COMMENTED{
Again, to keep notations simple, we consider just one operator
$\hat{B}$, the general case being a simple iteration of the
algorithm we will describe. 
}

We will exploit the following exact property
\begin{equation}
\label{smart}
 \hat{n}_{i,\sigma} = \frac{e^{\hat{n}_{i,\sigma}} - 1}{e-1}
\end{equation}
which can be easily proved by expanding the exponential operator and noting
$ \hat{n}^2_{i,\sigma} =  \hat{n}_{i,\sigma} $. 
Thus the numerator in Eq.~\eqref{dynamical estimator_new_gen_tb} can be viewed as propagating 
two Slater determinants.

\COMMENTED{
\begin{figure}[ptb]
\begin{center}
\includegraphics[width=6.0cm,angle=-90]{two-body-b-eps-converted-to.pdf}
\caption{(color online) The computation of two-particle imaginary-tune correlation functions with the new method.
The correlation $n(\tau)=\langle \Psi_0  | \, \hat{n}^{}_{i,\sigma} \, e^{-\tau  (\hat{H} - E_0) } \,  \hat{n}^{}_{i,\sigma'}  \, | \Psi_0 \rangle$ is calculated  for a $2 \times 1$ lattice hosting $1$ spin up
and $1$ spin down particles
at $U = -2$, and compared to exact diagonalization.}
\label{fig:benchmark2}
\end{center}
\end{figure}
}

If $| \phi_R \rangle = \hat{c}^{\dagger}_{|u_1,\uparrow\rangle} \dots \hat{c}^{\dagger}_{|u_{N_\uparrow},\uparrow\rangle}
 \hat{c}^{\dagger}_{|v_1,\downarrow\rangle} \dots \hat{c}^{\dagger}_{|v_{N_\downarrow},\downarrow\rangle} |0\rangle$,
\COMMENTED{
\begin{equation}
| \phi_R \rangle = \hat{c}^{\dagger}_{|u_1,\uparrow\rangle} \dots \hat{c}^{\dagger}_{|u_{N_\uparrow},\uparrow\rangle}
 \hat{c}^{\dagger}_{|v_1,\downarrow\rangle} \dots \hat{c}^{\dagger}_{|v_{N_\downarrow},\downarrow\rangle} |0\rangle\,,
\end{equation}
}
the identity in Eq.~\eqref{smart} provides the following (for example in the spin-up case):
\begin{equation}
 \hat{n}_{i,\uparrow}| \phi_R \rangle = \frac{ | \phi'_R(i) \rangle -  | \phi_R \rangle}{e-1}
\end{equation}
where:
\begin{equation}
| \phi_R'(i) \rangle =
\hat{c}^{\dagger}_{|e^{\hat{n}_{i,\uparrow}}u_1,\uparrow\rangle} \dots \hat{c}^{\dagger}_{|e^{\hat{n}_{i,\uparrow}}u_{N_\uparrow},\uparrow\rangle}
 \hat{c}^{\dagger}_{|v_1,\downarrow\rangle} \dots \hat{c}^{\dagger}_{|v_{N_\downarrow},\downarrow\rangle} |0\rangle\,.
\end{equation}
The application of the one-body propagator $e^{\hat{n}_{i,\uparrow}}$ on the orbitals can be carried out 
straightforwardly.
\COMMENTED{
with:

\begin{equation}
\langle j, \sigma \, |e^{\hat{n}_{i,\uparrow}}u_k,\uparrow\rangle = \delta_{\sigma,\uparrow}
\begin{cases}
& \langle j \, | u_k\rangle, \quad j \ne i \\
& e\,\, \langle j \, | u_k\rangle, \quad j = i
\end{cases}
\end{equation}
}
Now, the estimator Eq.~\eqref{dynamical estimator_new_gen_tb} can be broken into
two pieces:
\begin{equation}
\label{dynamical estimator_new_gen_tb-broken}
n({\bf{X}},\tau) = \frac{1}{e-1} \left(n_1({\bf{X}},\tau) - n_2({\bf{X}},\tau)\right)\,,
\end{equation}
which can be conveniently expressed as:
\begin{equation}
n_1({\bf{X}},\tau) = \frac{ \langle \phi_L  | \, \hat{n}_{j,\sigma'} \hat{B}  \,\,  | \phi'_R(i) \rangle }
{\langle \phi_L \, | \hat{B}  \,\,| \, \phi'_R (i)\rangle }
\,\, \frac{ \langle \phi_L  | \, \hat{B}  \,\,  | \phi'_R(i) \rangle }
{\langle \phi_L \, | \hat{B}  \,\,| \, \phi_R \rangle }
\end{equation}
and:
\begin{equation}
n_2({\bf{X}},\tau)  =  \frac{ \langle \phi_L  | \, \hat{n}_{j,\sigma'} \hat{B}  \,\,  | \phi_R \rangle }
{\langle \phi_L \, | \hat{B}  \,\,| \, \phi_R \rangle }
\end{equation}
Both $n_1({\bf{X}},\tau)$ and $n_2({\bf{X}},\tau)$ can be readily calculated.
As usual, GS decomposition can be applied periodically in the propagation of $ | \phi_R \rangle$
and  $| \phi'_R \rangle$ to ensure numerically stability.
\COMMENTED{
In Fig.~\ref{fig:benchmark2} we illustrate this method in a small system by a comparison with exact results. 
}

\COMMENTED{
and it is possible to perform the Gram-Schmidt decomposition on 
$| \, \phi_R \rangle$ and $| \, \phi'_R(i) \rangle $ to stabilize the 
numerical evaluations.

In Fig.~\ref{fig:benchmark2} we show a comparison with exact results
for a small system. We will publish results for two-particles
correlations in large systems in future works.
}

\section{Conclusions}
We study the calculation of dynamical 
correlation functions in imaginary time using auxiliary-field quantum
Monte Carlo. 
The use of twisted-boundary
conditions is systematically explored. 
One-body finite-size corrections help improve the convergence to the thermodynamic limit.
We study the role of special twists which correspond to the minimum corrected gap, and show
that this dramatically reduces finite-size effects.
In the repulsive Hubbard model at half-filling, the 
charge gaps and spectral functions are computed for different interaction strengths. 
Much higher accuracy was reached than previously possible, 
especially for small gap values.
We propose a new approach to compute the imaginary-time Green's functions by
explicitly varying the number of particles in the QMC random walk. 
This method has several advantages, including a much more favorable computational cost
in ``dilute'' systems where the size of the basis is significantly larger than the number of fermions 
(from cubic to linear in lattice/basis size). 

\section{Acknowledgments}

We thank Huy Nguyen for his contributions at early stages of this work. This work was supported by 
NSF (Grant No. DMR-1409510). E.V.~and S.~Z. were also supported by the  
Simons Foundation. Computing was carried out at the Extreme Science and Engineering Discovery Environment (XSEDE),
which is supported by National Science Foundation grant number ACI-1053575, and the computational
facilities at William and Mary.

\COMMENTED{

Focussing on situations where the famous sign problem
is not present, we have suggested a strategy to use twisted-boundary
conditions to dramatically reduce finite-size effects. We reached an 
accuracy level such that we can detect a non-zero charge gap in
the repulsive Hubbard model at half-filling in the weakly interacting
regime. Moreover, we have suggested a new approach to perform
the calculations in the dilute limit, which allows to reduce the computational
cost from cubic to linear in the lattice size, allowing to push forward
the calculations to huge systems: this can have major applications
in the physics of cold atoms.
}

%


\end{document}